\documentclass[11pt]{amsart}
\usepackage{latexsym,amssymb,amsmath,amscd,amsthm}
\numberwithin{equation}{section}
\theoremstyle{remark}
\newtheorem{theorem}{{\bf THEOREM}}[section]

\newtheorem{definition}{{\bf DEFINITION}}[section]

\newtheorem{example}{{\bf EXAMPLE}}[section]
\newtheorem{proposition}{{\bf PROPOSITION}}[section]
\newcommand{\bq}{\begin{equation}}
\newcommand{\bea}{\begin{array}}
\newcommand{\eea}{\end{array}}

\newcommand{\ga}{\alpha}
\newcommand{\gep}{\epsilon}
\newcommand{\gD}{\Delta}
\newcommand{\gl}{\lambda}
\newcommand{\gL}{\Lambda}
\newcommand{\gb}{\beta}

\newcommand{\mf}{\mathfrak}
\newcommand{\mc}{\mathcal}

\newcommand{\ul}[1]{\underline{#1}}
\newcommand{\ol}[1]{\overline{#1}}

\newcommand{\gO}{\Omega}
\newcommand{\gG}{\Gamma}
\newcommand{\gt}{\theta}
\newcommand{\gs}{\sigma}
\newcommand{\gz}{\zeta}
\newcommand{\gag}{\gamma}
\newcommand{\gd}{\delta}
\newcommand{\pp}{\partial}

\newcommand{\olra}{\overleftrightarrow}

\newcommand{\tl}{\tilde}
\newcommand{\na}{\nabla}

\newcommand{\bs}{\blacksquare}

\newcommand{\bgs}{\bigstar}

\newcommand{\gS}{\Sigma}

\newcommand{{\DDD}}{D\!\!\!\!\!\!-}

\newcommand{\bx}{\Box}

\newcommand{\emp}{\emptyset}
\title{ON THE QUANTUM POTENTIAL}
\author{Robert Carroll\\University of Illinois, Urbana, IL 61801}

\date{March, 2004\thanks{email: rcarroll@math.uiuc.edu}}

\begin{document}

\bibliographystyle{plain}

\begin{abstract}
We survey various origins and expressions for the quantum potential, expanding and extending
the treatment given in a previous paper \cite{c7}.
\end{abstract}

\maketitle

\tableofcontents

\section{INTRODUCTION}
\renewcommand{\theequation}{1.\arabic{equation}}
\setcounter{equation}{0}

The quantum potential arises in various forms, some of which were summarized in
\cite{c7}, and we want to return to this in a more systematic manner, with some
new embellishments.  Historically this arises by putting ${\bf (A1)}\,\,\psi=
Rexp(iS/\hbar)$ into the Schr\"odinger equation (SE) ${\bf (A2)}\,\,i\hbar\pp_t\psi
=-(\hbar^2/2m)\psi_{xx}+V\psi$ (1-D for simplicity), yielding 
\bq\label{1.1}
S_t+\frac{(S')^2}{2m}-\left(\frac{\hbar^2}{2m}\right)\left(\frac{R''}{R}\right)+V=0;
\,\,\pp_tR^2+\pp\left(\frac{R^2S'}{m}\right)=0
\end{equation}
The quantity ${\bf (A3)}\,\,Q=-(\hbar^2/2m)(R''/R)$ (or more generally $Q=
-(\hbar^2/2m)(\gD R/R)$) is the quantum potential and one takes ${\bf (A4)}\,\,
p=S'$ with $v=p/m$ (or $p=\na S$ with $v=p/m$) for the momentum and velocity.
We mention here in passing the refinements in \cite{b3,ch,c3,f2,f3,f7,f8} relative
to the stationary situation $S_t=-E$, which precludes the use of {\bf (A1)} as such
and leads to ${\bf (A5)}\,\,v=p/m_Q$ where $m_Q=m(1-\pp_EQ)$; this will be
discussed in more detail below.  In any event the quantum potential does enter into
any trajectory theory of deBroglie-Bohm (dBB) type.  The history is discussed 
for example in \cite{h99} (cf. also \cite{b2,b37,b38,b8,b7}) and we will show how
this quantum potential idea can be formulated in various ways in terms of
statistical mechanics, hydrodynamics, information and entropy, etc. when dealing
with different versions and origins of the SE.  Given the existence of particles
we finds the pilot wave of thinking very attractive, with the wave function serving
to choreograph the particle motion (or perhaps to ``create"  particles and/or
spacetime paths).
However the existence of particles itself is not such an assured matter and in
field theory approaches for example one will deal with particle currents
(cf. \cite{n57}  and see also e.g. \cite{b39,b40,d42,g1,t16}).  The whole idea of
quantum particle path seems in any case to be either fractal (cf. \cite{a11,a10,
a16,c7,c29,c20,c31,n11,n4,n13,n30,o2,o5,o6}, stochastic (see e.g.
\cite{b2,c7,f1,f10,h2,h3,h4,k8,k9,n9,n15,n6,r3}, or field theoretic (cf. 
\cite{b39,b40,d42,g1,n56,n57,n61,n62,n63,n64,n65,t16}.  The fractal approach sometimes
imagines an underlying micro-spacetime where paths are perhaps fractals with jumps,
etc. and one possible advantage of a field theoretic approach would be to let the fields
sense the ripples, which as e.g. operator valued Schwartz type distributions, they
could well accomplish. In fact what comes into question here is the structure of the
vacuum and/or of spacetime itself.  One can envision microstructures as in
\cite{c7,g95,n11,n12} for example, textures (topological defects) as in
\cite{b12,b13,w1}, Planck scale structure and QFT, along with space-time uncertainty
relations as in \cite {b11,d6,d7,l1,y1}, vacuum structures and conformal invariance as
in
\cite{m1,m2,s2,s4,s5,s6}, pilot wave cosmology as in \cite{s1}, ether theories as in
\cite{s21,s22}, etc.  Generally there seems to be a sense in which particles cannot be
measured as such and hence the idea of particle currents (perhaps corresponding to fuzzy
particles or ergodic clumps) should prevail perhaps along with the idea of probability
packets.  A number of arguments work with a (representative) trajectory as if it were a
single particle but there is no reason to take this too seriously; it could be thought of
perhaps as a ``typical" particle in a cloud but conclusions should perhaps always
be constructed from an ensemble point of view.  We will try to develop some of this
below.  The sticky point as we see it now goes as follows.  Even though one can
write stochastic equations for (typical) particle motion as in the Nelson theory
for example one runs into the problem of ever actually being able to localize a
particle.  Indeed as indicated in \cite{d6,d7} (working in a relativistic context
but this should hold in general) one expects space time uncertainty relations even
at a semiclassical level since any localization experiment will generate a
gravitational field and deform spacetime.  Thus there are relations 
$[q_{\mu},q_{\nu}]=i\gl^2_PQ_{\mu\nu}$ where $\gl_P$ is the Planck length and the
picture of spacetime as a local Minkowski manifold should break down at distances
of order $\gl_P$.  One wants the localization experiment to avoid creating a black
hole (putting the object out of ``reach") for example and this suggests 
$\gD x_0(\sum_1^3\gD x_i)\gtrsim\gl_P^2$ with $\gD x_1\gD x_2+\gD x_2\gD x_3+\gD
x_3\gD x_1\gtrsim\gl_P^2$ (cf. \cite{d6,d7}).  On the other hand in \cite{n57} it
is shown that in a relativistic bosonic field theory for example one can speak of
currents and n-particle wave functions can have particles attributed to them with
well defined trajectories, even though the probability of their experimental
detection is zero.  Thus one enters an arena of perfectly respectible but
undetectible particle trajectories.  The discussion in \cite{c64,d42,s52,v1,v2} is
also relevant here; some recourse to the idea of beables and reality and observables
as beables is also involved (cf. \cite{b39,b40,c64,v1}).  We will have something
to say about all these matters.

\section{POINTS OF VIEW}
\renewcommand{\theequation}{2.\arabic{equation}}
\setcounter{equation}{0}

We collect here some different ways in which the quantum potential arises with a
sketch of the derivation (cf. \cite{c7} for more details, derivations and references);
additional origins will be given subsequently.

\subsection{SCHR\"ODINGER EQUATIONS}

\begin{example}
Take {\bf (A1)}, {\bf (A2)}, and \eqref{1.1} with $P=R^2\,\,(\sim |\psi|^2)$ and Q as
in {\bf (A3)}.  This gives ${\bf (A6)}\,\,
S_t+\frac{(S')^2}{2m}+Q+V=0;\,\,P_t+\frac{1}{m}(PS')'=0$
which has some hydrodynamical interpretations in the spirit of Madelung.  Indeed  
going to \cite{d3}
for example we take $p=S'$ with $p=m\dot{q}$ for $\dot{q}$ a
velocity (or ``collective" velocity - unspecified).  Then {\bf (A6)} leads to
($\rho=mP$ is an unspecified mass density)
\bq\label{2.1}
P_t+(P\dot{q})'=0\equiv\rho_t+(\rho\dot{q})'=0;\,\,S_t+\frac{p^2}{2m}+V-
\frac{\hbar^2}{2m}\frac{\pp^2\sqrt{\rho}}{\sqrt{\rho}}=0
\end{equation}
A little calculation then yields
\bq\label{2.2}
\pp_t(\rho v)+\pp(\rho v^2)+\frac{\rho}{m}\pp V-\frac{\hbar^2}{2m^2}\rho\pp
\left(\frac{\pp^2\sqrt{\rho}}{\sqrt{\rho}}\right)=0
\end{equation}
This is similar to an ``Euler" type equation (cf. \cite{d3}) and it
definitely has a hydrodynamic flavor (cf. also \cite{g99}).
Now go to \cite{p4} and write from \eqref{2.1} 
\bq\label{2.3}
\frac{\pp v}{\pp t}+(v\cdot\na)v=-\frac{1}{m}\na(V+Q);\,\,v_t+vv'=-(1/m)\pp(v+Q)
\end{equation}
The higher dimensional form is not considered here but matters are similar there.
This equation (and \eqref{2.2}) is incomplete as a hydrodynamical equation as a
consequence of a missing term $-\rho^{-1}\na {\mf p}$ where ${\mf p}$ is the pressure
(cf. \cite{l5}).
Hence one ``completes" the equation in the form
\bq\label{2.4}
m\left(\frac{\pp v}{\pp t}+(v\cdot\na)v\right)=-\na(V+Q)-\na F;\,\,mv_t+mvv'=
-\pp(V+Q)-F'
\end{equation}
where ${\bf (A7)}\,\,\na F=(1/R^2)\na {\mf p}$ (or $F'=(1/R^2){\mf p}'$).  By the
derivations above this would then correspond to an extended SE of the form
${\bf (A8)}\,\,
i\hbar\frac{\pp\psi}{\pp t}=-\frac{\hbar^2}{2m}\gD\psi+V\psi+F\psi$
provided one can determine F in terms of the wave function $\psi$.  This suggests a
role for Q in the form ($F=0)\,\, {\bf (A9)}\,\,\pp Q=(1/\rho){\mf p}'\Rightarrow
{\mf p}'=-(\hbar^2/2m)\rho(\pp^2\sqrt{\rho}/\sqrt{\rho}=-(\hbar^2/2m)
R^2\pp(R''/R)$.  Alternatively one can form a nonlinear SE (NLSE) with ${\mf p}$ a
suitable function of $\psi$.$\hfill\bs$
\end{example}
\begin{example}
We turn next to \cite{k9} for a statistical origin for QM (cf. also
\cite{ch,c15,k8,k9,n6,o1,r6}).  The idea is to build a program in which the
microscopic motion, underlying QM, is described by a rigorous dynamics different
from Brownian motion (thus avoiding unnecessary assumptions about the Brownian
nature of the underlying dynamics).  The Madelung approach gives rise to fluid
dynamical type equations with a quantum potential, the latter being capable of
interpretation in terms of a stress tensor of a quantum fluid.
Thus one shows in \cite{k9} that the quantum state corresponds to a subquantum
statistical ensemble whose time evolution is governed by classical kinetics in
the phase space.  The equations take the form
\bq\label{2.5}
\rho_t+\pp_x(\rho u)=0;\,\,\pp_t(\mu\rho u_i)+\pp_j(\rho\phi_{ij})+\rho\pp_{x_i}V
=0;\,\,\pp_t(\rho E)+\pp_x(\rho S)-\rho\pp_tV=0
\end{equation}
with ${\bf (A10)}\,\,
\frac{\pp S}{\pp t}+\frac{1}{2\mu}\left(\frac{\pp S}{\pp x}\right)^2+{\mc W}+
V=0$
for two scalar fields $\rho,\,S$ determining a quantum fluid.  These can be rewritten
as
\bq\label{2.6}
\frac{\pp\xi}{\pp t}+\frac{1}{\mu}\frac{\pp^2S}{\pp x^2}+\frac{1}{\mu}\frac{\pp\xi}
{\pp x}\frac{\pp S}{\pp x}=0;
\end{equation}
$$\frac{\pp S}{\pp t}-\frac{\eta^2}{4\mu}\frac{\pp^2\xi}{\pp
x^2}-\frac{\eta^2}{8\mu}\left(\frac{\pp \xi}{\pp
x}\right)^2+\frac{1}{2\mu}\left(\frac{\pp S}{\pp x}\right)^2+V=0$$
where $\xi=log(\rho)$ and for $\gO=(\xi/2)+(i/\eta)S=log{\Psi}$ with $m=N\mu,\,\, 
{\mc V}=NV$, and $\hbar=N\eta$ one arrives at a SE
${\bf (A11)}\,\,
i\hbar\frac{\pp\Psi}{\pp t}=-\frac{\hbar^2}{2m}\frac{\pp^2 \Psi}{\pp x^2}+{\mc
V}\Psi$.
Further one can write $\Psi=\rho^{1/2}exp(i{\mf S}/\hbar)$ with ${\mf S}=NS$ and here
$N=\int|\Psi|^2d^nx$.  Thus from a statistical origin in classical kinetics there
emerges a SE with quantum potential ${\mc W}\sim Q$ (before scaling with N) related to
the stress tensor of a quantum fluid.
$\hfill\bs$
\end{example}
\begin{example}
Now in \cite{d4} one is obliged to use the form $\psi=Rexp(iS/\hbar)$ to make sense
out of the constructions (this is no problem with suitable provisos, e.g. that S is not
constant - cf. \cite{b3,ch,f2,f3} and comments later).  This leads to \eqref{1.1}
and \eqref{2.1} with 
$Q=-\hbar^2R''/2mR$ as in {\bf (A3)}.
In \cite{d4} one emphasizes 
configurations based on coordinates whose motion is choreographed by the SE according
to the rule (1-D only here)
\bq\label{2.7}
\dot{q}=v=\frac{\hbar}{m}\Im\frac{\psi^*\psi'}{|\psi|^2}=\frac{\hbar}{m}\Im\left(\frac
{\psi'}{\psi}\right)
\end{equation}
where ${\bf (A12)}\,\,i\hbar\psi_t=-(\hbar^2/2m)\psi''+V\psi$.  The argument for 
\eqref{2.7} is based on obtaining the simplest Galilean and time reversal invariant
form for velocity, transforming correctly under velocity boosts.  This leads directly
to \eqref{2.7} so that Bohmian mechanics (BM) is governed by
\eqref{2.7} and {\bf (A12)}.  It's a fairly convincing argument and no recourse to
Floydian time seems possible (cf. \cite{ch,f3,f7,f8}).  Note however that if $S=c$ then
$\dot{q} =v=(\hbar/m)\Im(R'/R)=0$ while $p=S'=0$ so perhaps this formulation avoids
the $S=0$ problems indicated in \cite{ch,f3,f7,f8}.
What makes the constant
$\hbar/m$ in \eqref{2.7} important here is that with this value the probability
density
$|\psi|^2$ on configuration space is equivariant.  This means that via the evolution
of probability densities $\rho_t+div(v\rho)=0$ (as in \eqref{2.1} with $v\sim p/m$)
the density $\rho=|\psi|^2$ is stationary relative to $\psi$, i.e. $\rho(t)$ retains
the form $|\psi(q,t)|^2$.  One calls $\rho=|\psi|^2$ the quantum equilibrium density
(QED) and says that a system is in quantum equilibrium when its coordinates are
randomly distributed according to the QED.  The quantum equilibrium hypothesis
(QHP) is the assertion that when a system has wave function $\psi$ the distribution
$\rho$ of its coordinates satisfies $\rho=|\psi|^2$.$\hfill\bs$
\end{example}
\begin{example}
The Fisher information connection (cf. \cite{h2,h3,h4}) involves a classical ensemble
with particle mass m moving under a potential V
${\bf (A13)}\,\,
S_t+\frac{1}{2m}(S')^2+V=0;\,\,P_t+\frac{1}{m}\pp(PS')'=0$
where S is a momentum potential; note that no quantum potential is present but this will
be added on in the form of a term $(1/2m)\int dt(\gD N)^2$ in the Lagrangian which measures
the strength of fluctuations.  This can then be specified in terms of the probability
density P leading to a SE.  A ``neater" approach
is given in following \cite{r3} leading in 1-D to
\bq\label{2.8}
S_t+\frac{1}{2m}(S')^2+V+\frac{\gl}{m}\left(\frac{(P')^2}{P^2}-\frac{2P''}{P}\right)=0
\end{equation}
Note that $Q=-(\hbar^2/2m)(R''/R)$ becomes for $R=P^{1/2}$ ${\bf (A14)}\,\,
Q=-(2\hbar^2/2m)[(2P''/P)-(P'/P)^2]$.  Thus the addition of the
Fisher information serves to quantize the classical system.
One also defines an information entropy (IE) via ${\bf (A15)}\,\,{\mf S}=-\int
\rho log(\rho)d^3x\,\,(\rho=|\psi|^2)$ leading to 
\bq\label{2.9}
\frac{\pp{\mf S}}{\pp t}=\int (1+log(\rho))\pp(v\rho)\sim \int \frac{(\rho')^2}{\rho}
\end{equation}
modulo constants involving $D\sim \hbar/2m$.  ${\mf S}$ is typically not conserved and 
$\pp_t\rho=-\na\cdot(v\rho)\,\,(u=D\na log(\rho)$ with $v=-u$ corresponds to standard
Brownian motion with
$d{\mf S}/dt\geq 0$.  Then high IE production corresponds to rapid flattening of the
probability density.  Note here also that ${\mf F}\sim -(2/D^2)\int \rho Qdx=\int
dx[(\rho')^2/\rho]$ is a functional form of Fisher information.  Entropy balance is
discussed in \cite{g10}.$\hfill\bs$
\end{example}
\indent
The Nagasawa theory (based in part on Nelson's approach) is very revealing
and fascinating (cf. \cite{n9,n15} and for the nonrelativistic theory one has
\begin{theorem}
Let $\psi(t,x)=exp[R(t,x)+iS(t,x)]$ be a solution of the SE ${\bf (A16)}\,\,
i\pp_t\psi+(1/2)\gD\psi+ia(t,x)\cdot\na\psi-V(t,x)\psi=0$ 
($\hbar$ and $m$ omitted) then
${\bf (A17)}\,\,\phi(t,x)=exp[R(t,x)+S(t,x)]$ and $\hat{\phi}=exp[R(t,x)-S(t,x)]$ are
solutions of 
\bq\label{2.10}
\frac{\pp\phi}{\pp t}+\frac{1}{2}\gD\phi+a(t,x)\cdot\na\phi+c(t,x,\phi)\phi=0;
\end{equation}
$$-\frac{\pp\hat{\phi}}{\pp t}+\frac{1}{2}\gD\hat{\phi}-a(t,x)\cdot\na\hat{\phi}
+c(t,x,\phi)\hat{\phi}=0$$
where the creation and annihilation term $c(t,x,\phi)$ is given via
\bq\label{2.11}
c(t,x,\phi)=-V(t,x)-2\frac{\pp S}{\pp t}(t,x)-(\na S)^2(t,x)-2a\cdot\na S(t,x)
\end{equation}
Conversely given $(\phi,\hat{\phi})$ as in {\bf (A17)} satisfying \eqref{2.10}
it follows that $\psi$ satisfies the SE {\bf (A16)}
with V as in
\eqref{2.11} (note $R=(1/2)log(\hat{\phi}\phi),\,\,
S=(1/2)log(\phi/\hat{\phi}),$ and $exp(R)=(\hat{\phi}\phi)^{1/2}$).
$\hfill\bs$
\end{theorem}
\indent
Thus in short: $\psi=exp(R+iS)$ satisfies the SE
${\bf (A18)}\,\,i\psi_t+(1/2)\psi''+ia\psi'-V\psi=0$ if and only if
\bq\label{2.12}
V=-S_t+\frac{1}{2}R''+\frac{1}{2}(R')^2-\frac{1}{2}(S')^2-aS;\,\,0=R_t+\frac{1}{2}
S''+S'R'+aR'
\end{equation}
Changing variables via $X=(\hbar/\sqrt{m})x$ and $T=\hbar t$ one gets
${\bf (A19)}\,\,i\hbar\psi_T=-(\hbar^2/2m)\psi_{XX}-iA\psi_X+V\psi$ where
$A=a\hbar/\sqrt{m}$  and 
\bq\label{2.13}
i\hbar R_T+(\hbar^2/m^2)R_XS_X+(\hbar^2/2m^2)S_{XX}+AR_X=0;
\end{equation} 
$$V=-i\hbar S_T+(\hbar^2/2m)R_{XX}+(\hbar^2/2m^2)R_X^2-(\hbar^2/2m^2)S_X^2-AS_X$$
The diffusion equations then take the form
\bq\label{2.14}
\hbar\phi_T+\frac{\hbar^2}{2m}\phi_{XX}+A\phi_X+\tl{c}\phi=0;\,\,-\hbar\hat{\phi}_T+
\frac{\hbar^2}{2m}\hat{\phi}_{XX}-A\hat{\phi}_X+\tl{c}\hat{\phi}=0;
\end{equation}
$$\tl{c}=-\tl{V}(X,T)-2\hbar S_T-\frac{\hbar^2}{m}S_X^2-2AS_X$$
It is now possible to introduce a role for the quantum potential in this theory.  Thus
from $\psi=exp(R+iS)$ (with $\hbar=m=1$ say) we have $\psi=\rho^{1/2}exp(iS)$ with
$\rho^{1/2}=exp(R)$ or $R=(1/2)log(\rho)$.  Hence $(1/2)(\rho'/\rho)=R'$ and $R''=(1/2)
[(\rho''/\rho)-(\rho'/\rho)^2]$ while the quantum potential is
$Q=(1/2)(\pp^2\rho^{1/2}/\rho^{1/2})=-(1/8)[(2\rho''/\rho)-(\rho'/\rho)^2]$ (cf. 
\eqref{2.4}).  Equation \eqref{2.12} becomes then
\bq\label{2.15}
V=-S_t+\frac{1}{8}\left(\frac{2\rho''}{\rho}-\frac{(\rho')^2}{\rho^2}\right)-\frac{1}{2}
(S')^2-aS
\equiv S_t+\frac{1}{2}(S')^2+V+Q+aS=0;
\end{equation}
$$\rho_t+\rho S''+S'\rho'+a\rho'=0\equiv \rho_t+(\rho S')'+a\rho'=0$$
Thus $-2S_t-(S')^2=2V+2Q+2aS$ and one has
\begin{proposition}
The creation-annihilation term $c$ in the diffusion equations (cf. Theorem 2.1) becomes
${\bf (A20)}\,\,
c=-V-2S_t-(S')^2-2aS'=V+2Q+2a(S-S')$
where Q is the quantum potential.$\hfill\bs$
\end{proposition}
\indent
The relativistic theory involves Markov processes with jumps for which we refer to \cite{n15}
(cf. also Example 3.4).
We remark also that the papers in \cite{d99} contain very interesting derivations of
Schr\"odinger equations via diffusion ideas \`a la Nelson, Markov wave equations, and
suitable ``applied" forces (e.g. radiative reactive forces).
\begin{example}
Regarding scale relativity one writes (cf. also Section 4) 
\bq\label{2.16}
\frac{d_{\pm}}{dt}x(t)=lim_{\gD t\to 0_{\pm}}\left<\frac{\pm y(t\pm\gD t)\mp y(t)}{\gD t}
\right>=b_{\pm}(t)
\end{equation}
and for $f=f(x(t),t)$ (smooth in $x$) we collect equations in ($\rho=|\psi|^2$)
\bq\label{2.17}
dx=b_{+}dt+d\xi_{+}=b_{-}dt+d\xi_{-};<d\xi_{+}^2>=2{\mc D}dt=-<d\xi_{-}^2>
\end{equation}
\bq\label{2.18}
\frac{d_{+}f}{dt}=(\pp_t+b_{+}\pp+{\mc
D}\pp^2)f;\,\,\frac{d_{-}f}{dt}=(\pp_t+b_{-}\pp-{\mc D}\pp^2)f
\end{equation}
\bq\label{2.19}
V=\frac{1}{2}(b_{+}+b_{-});\,\,U=\frac{1}{2}(b_{+}-b_{-});\,\,\rho_t+\pp(\rho
V)=0;\,\,U={\mc D}\pp(log(\rho));
\end{equation}
$${\mc V}=V-iU;\,\,d_{{\mc V}}=\frac{1}{2}(d_{+}+d_{-});\,\,d_{{\mc
U}}=\frac{1}{2}(d_{+}-d_{-})$$
\bq\label{2.20}
\frac{d_{{\mc V}}}{dt}=\pp_t+V\pp;\,\,\frac{d_{{\mc U}}}{dt}={\mc D}\pp^2+U\pp;\,\,\frac
{d'}{dt}=(\pp_t-i{\mc D}\pp^2)+{\mc V}\pp
\end{equation}
\bq\label{2.21}
V=2{\mc D}\pp S;\,\,{\mc S}=log(\rho^{1/2})+iS;\,\,\psi=\sqrt{\rho}e^{iS}=e^{i{\mc
S}};\,\,{\mc V}=-2i{\mc D}\pp log(\psi)
\end{equation}
For Lagrangian ${\mc L}=(1/2)m{\mc V}^2-m{\mf U}$ one gets a SE
${\bf (A21)}\,\,
i\hbar\psi_t=-\frac{\hbar^2}{2m}\pp^2\psi+{\mf U}\psi$
coming from Newton's law ${\bf (A22)}\,\,-\pp{\mf U}=-2i{\mc D}m(d'/dt)\pp
log(\psi)=m(d'/dt){\mc V}$.
If ${\mf U}=0$ we see that free motion $m(d'/dt){\mc
V}=0$ yields the SE as a geodesic equation in fractal spacetime.  To clarify this we
write out $(d'/dt)\pp log(\psi)=0$ as $(\pp_t-i{\mc D}\pp^2+V\pp)\pp log(\psi)=0$
with $V=-2i{\mc D}\pp log(\psi)$.  Using identities as in \cite{n5} this is
\bq\label{A1}
\pp_t\pp log(\psi)-i{\mc D}\pp^3log(\psi)-2i{\mc D}\pp log(\psi)\pp log(\psi)=0
\Rightarrow i\pp_t\psi+{\mc D}\pp^2\psi=0
\end{equation}
with ${\mc D}=\hbar/2m$ which means $i\hbar\pp_t\psi=-(\hbar^2/2m)\pp^2\psi$ as
desired (the multidimensional case is in \cite{n5}).  Next since the quantum
potential in this context arises via $Q=-(\hbar^2/2m)(\gD\sqrt{\rho}/\sqrt{\rho})$
(as usual) and since ${\mc V}=-(i\hbar/m)\pp
log(\psi)=V-iU=-(\hbar/m)\pp[log\sqrt{\rho}+iS]$ one can write
$(\bgs)\,\,V=(\hbar/m)\pp S$ and $U=(\hbar/m)(\pp\sqrt{\rho}/\sqrt{\rho})$.
Consequently
\bq\label{ugh}
\pp U=-\frac{\hbar}{m}\left(\frac{2m}{\hbar^2}Q+\frac{m^2}{\hbar^2}U^2\right)
\end{equation}
and $(\bgs\bgs)\,\,\pp U=-(2/\hbar)Q-(m\hbar)U^2\Rightarrow Q=-(m/2)U^2-(\hbar/2)
\pp U$ (cf. {\bf (A23)} below).  Hence the quantum potential arises directly from the
geodesic equation in fractal spacetime based on continuous nonsmooth paths.  We refer here
to Remark 5.2 for more discussion on the derivation of the SE.
$\hfill\bs$
\end{example}\indent
{\bf REMARK 2.1.}
It is perhaps a little unsettling to see operators $\pp$ and $\pp^2$ appearing in the
equations above since the paths $y(t)$ are not smooth.  However the functions 
$f(x(t),t)$ can well be smooth functions of x so there is less of a problem (see e.g.
\cite{a10,c7,c13,c31,n5} and Section 5.1 - the development of Cresson et al is rigorous and
polished).  For a sketch of explanation we follow \cite{c31} and consider $x\to f(x(t),t)\in
C^{n+1}$ with $X(t)\in H^{1/n}$ (i.e. $c\gep^{1/n}\leq |X(t')-X(t)|\leq C\gep^{1/n}$).
Define ($f$ real valued)
\bq\label{star}
\na_{\pm}^{\gep}f(t)=\frac{f(t\pm \gep)-f(t)}{\pm \gep};\,\,\frac{\bx_{\gep}f}{\bx t}
(f)=\frac{1}{2}(\na_{+}^{\gep}+\na_{-}^{\gep})f-\frac{i}{2}(\na_{+}^{\gep}-
\na_{-}^{\gep}f);
\end{equation}
$$a_{\gep,j}(t)=\frac{1}{2}[(\gD_{+}^{\gep}y)^j-(-1)^j(\gD_{-}^{\gep}y)^j]
-\frac{i}{2}[(\gD_{+}^{\gep}y)^j+(-1)^j(\gD_{-}^{\gep}y)^j]$$
Assume some minimal control over the lack of differentiability (cf. \cite{c31}) and
then for $f$ now complex valued with $\bx_{\gep}f/\bx t=(\bx_{\gep}f_{\Re}/\bx t)+i
(\bx_{\gep}f_{\Im}/\bx t)$ (note the mixing of i terms is not trivial) one has
\bq\label{starstar}
\frac{\bx_{\gep}f}{\bx t}=\frac{\pp f}{\pp t}+\frac{\bx_{\gep}x}{\bx t}\frac{\pp
f}{\pp x}+\sum_2^n\frac{1}{j!}a_{\gep,j}(t)\frac{\pp^jf}{\pp
x^j}\gep^{j-1}+o(\gep^{1/n})
\end{equation}
We refer to \cite{a10,c13,c31} for a full exposition.$\hfill\bs$
\\[3mm]\indent
\begin{example}
The development in \cite{c29} involves thinking of nonlinear QM as a
fractal Brownian motion with complex diffusion coefficient.  In particular one uses
{\bf (A22)} and \eqref{2.20} and arrives at
\bq\label{2.23}
-\na U=-2im[D\pp_t\na log(\psi)]-2D\na\left(D\frac{\na^2\psi}{\psi}\right)
\end{equation}
Thus putting in a complex diffusion coefficient leads to the NLSE
\bq\label{2.24}
i\hbar\pp_t\psi=-\frac{\hbar^2}{2m}\frac{\ga}{\hbar}\na^2\psi+U\psi-i\frac{\hbar^2}{2m}
\frac{\gb}{\hbar}(\na log(\psi))^2\psi
\end{equation}
with $\hbar=\ga+i\gb=2mD$ complex.
\\[3mm]\indent
In \cite{a16} one writes again $\psi=Rexp(iS/\hbar)$ with field equations in the
hydrodynamical picture
\bq\label{2.25}
d_t(m_0\rho v)=\pp_t(m_0\rho v)+\na(m_0\rho v)=-\rho\na(u+Q);\,\,\pp_t\rho+\na\cdot(\rho v)=0
\end{equation}
where $Q=-(\hbar^2/2m_0)(\gD\sqrt{\rho}/\sqrt{\rho})$.  One works with the Nottale approach
as above with $d_v\sim d_{{\mc V}}$ and $d_u\sim d_{{\mc U}}$ (cf. \eqref{2.20}).  One
assumes that the velocity field from the hydrodynamical model agrees with the real part
$v$ of the complex velocity $V=v-iu$ so (cf. \eqref{2.19}) $v=(1/m_0)\na s\sim 2D\pp s$
and $u=-(1/m_0)
\na\gs\sim D\pp log(\rho)$ where $D=\hbar/2m_0$.  In this context the quantum potential
$Q=-(\hbar^2/2m_0)\gD\sqrt{\rho}/\sqrt{\rho}$ becomes ${\bf (A23)}\,\,Q=-m_0D\na
u-(1/2)m_0u^2\sim -(\hbar/2)\pp u-(1/2)m_0u^2$.  Consequently Q arises from the fractal
derivative and the nondifferentiability of spacetime.  Further one can relate $u$ (and hence
Q) to an internal stress tensor ${\bf (A24)}\,\,\gs_{ik}=\eta[(\pp u_i/\pp x_k)+
(\pp u_k/\pp x_i)]$
whereas the $v$ equations correspond to
systems of Navier-Stokes type.$\hfill\bs$ 
\end{example}
\begin{example}
The equivalence principle (EP) of Faraggi-Matone (cf. \cite{b3,ch,c2,c9,f3}) is based on
the idea that all physical systems can be connected by a coordinate transformation to
the free situation with vanishing energy (i.e. all potentials are equivalent under
coordinate transformations).  This automatically leads to the quantum stationary
Hamilton-Jacobi equation (QSHJE) which is a third order nonlinear differential
equation providing a trajectory representation of quantum mechanics (QM).  The theory
transcends in several respects the Bohm theory and in particular utilizes a Floydian
time (cf. \cite{f7,f8})
leading to ${\bf (A25)}\,\,\dot{q}=p/m_Q\ne p/m$ where ${\bf
(A26)}\,\,m_Q=m(1-\partial_EQ)$ is the ``quantum mass" and Q the ``quantum potential". 
Thus the EP is reminscient of the Einstein equivalence of relativity theory.  This
latter served as a midwife to the birth of relativity but was somewhat inaccurate in its
original form.  It is better put as saying that all laws of physics should be invariant
under general coordinate transformations (cf. \cite{o9}).  This demands that not only
the form but also the content of the equations be unchanged.  More precisely the
equations should be covariant and all absolute constants in the equations are to be left
unchanged (e.g. $c,\,\hbar,\,e,\,m$ and $\eta_{\mu\nu}=$ Minkowski tensor).  
Now for the EP, the classical picture with $S^{cl}(q,Q^0,t)$ the Hamilton
principal function ($p=\partial S^{cl}/\partial q$) and $P^0,\,Q^0$ playing the role of
initial conditions involves the classical HJ equation (CHJE) ${\bf (A27)}\,\,H(q,p=
(\partial S^{cl}/\partial q),t)+(\partial S^{cl}/\partial t)=0$.  For time independent V one
writes
$S^{cl}=S_0^{cl}(q,Q^0)-Et$ and arrives at the classical stationary HJ equation
(CSHJE) ${\bf (A28)}\,\,(1/2m)(\partial S_0^{cl}/\partial q)^2+{\mathfrak W}=0$ where
${\mathfrak W}=V(q)-E$.  In the Bohm theory one looked at Schr\"odinger equations ${\bf
(A29)}\,\, i\hbar\psi_t=-(\hbar^2/2m)\psi''+V\psi$ with $\psi=\psi(q)exp(-iEt/\hbar)$
and
${\bf (A30)}\,\,\psi(q)=R(qexp(i\hat{W}/\hbar)$ leading to 
\bq\label{2.25}
\left(\frac{1}{2m}\right)(\hat{W}')^2+V-E-\frac{\hbar^2R''}{2mR}=0;\,\,
(R^2\hat{W}')'=0
\end{equation}
where ${\bf (A31)}\,\,\hat{Q}=-\hbar^2R''/2mR$ was called the quantum potential; this
can be written in the Schwartzian form ${\bf
(A32)}\,\,\hat{Q}=(\hbar^2/4m)\{\hat{W};q\}$ (via $R^2\hat{W}'=c$).  Here ${\bf
(A33)}\,\,\{f;q\}=(f'''/f')-(3/2)(f''/f')^2$.  Writing ${\mathfrak W}=V(q)-E$ as in 
{\bf (A28)} we have the quantum stationary HJ equation (QSHJE)
${\bf (A34)}\,\,(1/2m)(\partial\hat{W}'/\partial q)^2+{\mathfrak W}(q)+\hat{Q}(q)=0$
($\equiv {\mathfrak W}=-(\hbar^2/4m)\{exp(2iS_0/\hbar);q\}$).
This was 
worked out in the Bohm school (without the Schwarzian connections) but {\bf (A30)}
is not appropriate for all situations; the Bohm theory is incomplete and can lead to
incorrect predictions ($\hat{W}=constant$ must be excluded).
The technique of Faraggi-Matone (FM) is completely general and with only the EP as
guide one exploits the relations between Schwarzians, Legendre duality, and the
geometry of a second order differential operator $D_x^2+V(x)$ (M\"obius
transformations play an important role here) to arrive at the QSHJE in the form
\bq\label{2.26}
\frac{1}{2m}\left(\frac{\partial S_0^v(q^v)}{\partial q^v}\right)^2+{\mathfrak W}(q^v)+
{\mathfrak Q}^v(q^v)=0
\end{equation}
where $v:\,q\to q^v$ represents an arbitrary locally invertible coordinate
transformation.  Note in this direction for example that the Schwarzian derivative
of the the ratio of two linearly independent elements in $ker(D^2_x+V(x))$ is twice
$V(x)$.  In particular given an arbitrary system with coordinate $q$ and reduced
action $S_0(q)$ the system with coordinate $q^0$ corresponding to $V-E=0$ involves 
${\bf (A35)}\,\,{\mathfrak W}(q)=(q^0;q)$ where $(q^0,q)$ is a cocycle term which has
the form ${\bf (A36)}\,\,(q^a;q^b)=-(\hbar^2/4m)\{q^a;q^b\}$.  In fact it can be said
that the essence of the EP is the cocycle conditon ${\bf
(A37)}\,\,(q^a;q^c)=(\partial_{q^c}q^b)^2[(q^a;q^b)-(q^c;q^b)]$.
\\[3mm]\indent
In addition FM developed a theory of $(x,\psi)$ duality (cf. ref. [1]) which
related the space coordinate and the wave function via a prepotential (free energy)
in the form ${\mathfrak F}=(1/2)\psi\bar{\psi}+iX/\epsilon$ for example.
A number of interesting philosophical points arise (e.g. the emergence of space from
the wave function) and we connected this to various features of
dispersionless KdV in \cite{ch,c9} in a sort of extended WKB spirit.  One should
note here that although a form {\bf (A30)} is not generally appropriate it is correct
when one is dealing with two independent solutions of the Schr\"odinger equation
$\psi$ and $\bar{\psi}$ which are not proportional.  In this context we utilized
some interplay between various geometric properties of KdV which involve the Lax
operator $L^2=D_x^2+V(x)$ and of course this is all related to Schwartzians,
Virasoro algebras, and vector fields on $S^1$ (see e.g. \cite{ch,c37,c38,c39,c40}).
Thus the simple presence of the
Schr\"odinger equation (SE) in QM automatically incorporates a host of geometrical
properties of $D_x=d/dx$ and the circle $S^1$.  In fact since the FM theory exhibits the
fundamental nature of the  SE via its geometrical properties connected to the QSHJE one
could speculate about trivializing QM (for 1-D) to a study of $S^1$ and $\partial_x$.
$\hfill\bs$
\end{example}

\subsection{THE KLEIN-GORDON EQUATION}

We import here some comments based on \cite{b3} concerning the Klein-Gordon (KG)
equation and the equivalence principle (EP) of Example 2.7 
(details are in \cite{b3} and cf. also \cite{d20,d17,h99,m5,m6} for the KG equation).
One starts with the relativistic classical Hamilton-Jacobi equation (RCHJE) with a
potential $V(q,t)$ given as 
\bq\label{2.27}
\frac{1}{2m}\sum_1^D(\pp_kS^{cl}(q,t))^2+{\mf W}_{rel}(q,t)=0;\,\,{\mf W}_{rel}(q,t)=
\frac{1}{2mc^2}[m^2c^4-(V(q,t)+\pp_tS^{cl}(q,t))^2]
\end{equation}
In the time-independent case one has $S^{cl}(q,t)=S_0^{cl}(q)-Et$ and \eqref{2.26}
becomes 
\bq\label{2.28}
\frac{1}{2m}\sum_1^D(\pp_kS_0^{cl})^2+{\mf W}_{rel}=0;\,\,{\mf W}_{rel}(q)=
\frac{1}{2mc^2}[m^2c^4-(V(q)-E)^2]
\end{equation}
In the latter case one can go through the same steps as in the nonrelativistic case and 
the relativistic quantum HJ equation (RQHJE) becomes ${\bf (A38)}\,\,(1/2m)(\na
S_0)^2+{\mf W}_{rel}-(\hbar^2/2m)(\gD R/R)=0$ with $\na\cdot(R^2\na S_0)=0$; these
equations imply the stationary KG equation ${\bf (A39)}\,\,-\hbar^2c^2\gD
\psi+(m^2c^4-V^2+2EV-E^2)\psi=0$ where
$\psi=Rexp(iS_0/\hbar)$.  Now in the time dependent case the (D+1)-dimensional RCHJE
is ${\bf (A40)}\,\,(1/2m)\eta^{\mu\nu}\pp_{\mu}S^{cl}\pp_{\nu}S^{cl}
+{\mf W}'_{rel}=0$ where $\eta^{\mu\nu}=diag(-1,1,\cdots,1)$
and ${\bf (A41)}\,\,{\mf W}'_{rel}=(1/2mc^2)[m^2c^4-V^2(q)-2cV(q)\pp_0S^{cl}(q)]$
with $q=(ct,q_1,\cdots,q_D)$.  Thus {\bf (A40)} has the same structure as \eqref{2.28}
with Euclidean metric replaced by the Minkowskian one.  We know how to implement the EP
by adding Q via ${\bf (A42)}\,\,(1/2m)(\pp S)^2+{\mf W}_{rel}+Q=0$ (cf. \cite{f3} and
Example 2.7).  Note now that ${\mf W}'_{rel}$ depends on $S^{cl}$ one requires an
identification
${\bf (A43)}\,\,{\mf W}_{rel}=(1/2mc^2)[m^2c^4-V^2(q)-2cV(q)\pp_0S(q)]$ 
($S$ replacing $S^{cl}$) and
implementation of the EP requires that for an arbitrary ${\mf W}^a$ state ($q\sim q^a$)
one must have ${\bf (A44)}\,\,{\mf W}_{rel}^b(q^b)=(p^b|p^a){\mf W}_{rel}^a(q^a)
+(q^q;q^b)$ (cf. {\bf (A36)}) and ${\bf (A45)}\,\,Q^b(q^b)=(p^b|p^a)Q(q^a)-(q^a;q^b)$
where ${\bf
(A46)}\,\,(p^b|p)=[\eta^{\mu\nu}p_{\mu}^bp_{\nu}^b/\eta^{\mu\nu}p_{\mu}p_{\nu}]=
p^TJ\eta J^Tp/p^T\eta p$ and ${\bf (A47)}\,\,J_{\nu}^{\mu}=\pp q^{\mu}/\pp q^{b^{\nu}}$
(J is a Jacobian and these formulas are the natural multidimensional generalization
- see \cite{b3} for details).  Furthermore there is a cocycle condition ${\bf (A48)}\,\,
(q^a;q^c)=(p^c|p^b)[(q^a;q^b)-(q^c;q^b)]$ (cf. {\bf (A37)}).
\\[3mm]\indent
Next one shows that ${\bf (A49)}\,\,{\mf W}_{rel}=(\hbar^2/2m)[\bx
(Rexp(iS/\hbar))/Rexp(iS/\hbar)]$ and hence {\bf the corresponding quantum potential
is ${\bf (A50)}\,\,Q_{rel}=-(\hbar^2/2m) [\bx R/R]$.}  Then the RQHJE becomes ${\bf
(A51)}\,\,(1/2m)(\pp S)^2+{\mf W}_{rel} +Q=0$ with $\pp\cdot (R^2\pp S)=0$ (here $\bx
R=\pp_{\mu}\pp^{\mu}R$) and this reduces to the standard SE in the classical limit $c\to
\infty$.  To see how the EP is simply implemented one considers the so called minimal
coupling prescription for an interaction with an electromagnetic four vector
$A_{\mu}$.  Thus set
$P_{\mu}^{cl}=p_{\mu}^{cl}+eA_{\mu}$ where $p_{\mu}^{cl}$ is a particle momentum and
$P_{\mu}^{cl}=\pp_{\mu} S^{cl}$ is the generalized momentum.  Then the RCHJE reads
as ${\bf (A52)}\,\,(1/2m)(\pp S^{cl}-eA)^2+(1/2)mc^2=0$ where $A_0=-V/ec$.  Then
${\bf (A53)}\,\,{\mf W}=(1/2)mc^2$ and the critical case ${\mf W}=0$ corresponds to the
limit situation where $m=0$.  One adds the standard Q correction for implementation of
the EP to get
${\bf (A54)}\,\,(1/2m)(\pp S-eA)^2+(1/2)mc^2+Q=0$ and there are transformation
properties
\bq\label{2.29}
{\mf W}(q^b)=(p^b|p^a){\mf W}^a(q^a)+(q^a;q^b);\,\,Q^b(q^b)=(p^q|p^a)Q^a(q^a)-
(q^a;q^b)
\end{equation}
$$(p^b|p)=\frac{(p^b-eA^b)^2}{(p-eA)^2}=\frac{(p-eA)^TJ\eta J^T(p-eA)}{(p-eA)^T
\eta (p-eA)}$$
Here J is a Jacobian ${\bf (A55)}\,\,J_{\nu}^{\mu}=\pp q^{\mu}/\pp q^{b^{\nu}}$ and
this all implies the cocycle condition {\bf (A37)} again.  One finds now that
\bq\label{2.30}
(\pp S-eA)^2=\hbar^2\left(\frac{\bx R}{R}-\frac{D^2(Re^{iS/\hbar})}
{Re^{iS/\hbar}}\right);\,\,D_{\mu}=\pp_{\mu}-\frac{i}{\hbar}eA_{\mu}
\end{equation}
and it follows that
\bq\label{2.31}
{\mf W}=\frac{\hbar^2}{2m}\frac{D^2(Re^{iS/\hbar})}{Re^{iS/\hbar}};\,\,Q=-\frac
{\hbar^2}{2m}\frac{\bx R}{R};\,\,D^2=\bx-\frac{2ieA\pp}{\hbar}-\frac{e^2A^2}{\hbar^2}
-\frac{ie\pp A}{\hbar}
\end{equation}
\bq\label{2.32}
(\pp S-eA)^2+m^2c^2-\hbar^2\frac{\bx R}{R}=0;\,\,\pp\cdot (R^2(\pp S-eA))=0
\end{equation}
Note also that {\bf (A40)} coincides with {\bf (A52)} after setting ${\mf W}_{rel}
=mc^2/2$ and replacing $\pp_{\mu}S^{cl}$ by $\pp_{\mu}S^{cl}-eA_{\mu}$.  One can check
that \eqref{2.32} implies the KG equation ${\bf (A56)}\,\,(i\hbar\pp+eA)^2\psi+
m^2c^2\psi=0$ with $\psi=Rexp(iS/\hbar)$.  
\\[3mm]\indent
{\bf REMARK 2.2.}
We extract now a remark about mass generation and the EP from \cite{b3}.  Thus a
special property of the EP is that it cannot be implemented in classical mechanics
(CM) because of the fixed point corresponding to ${\mf W}=0$.  One is forced to
introduce a uniquely determined piece to the classical HJ equation (namely a quantum
potential Q). In the case of the RCHJE {\bf (A52)} the fixed point ${\mf W}(q^0)=0$
corresponds to
$m=0$ and the EP then implies that all the other masses can be generated by a
coordinate transformation.  Consequently one concludes that masses correspond to
the inhomogeneous term in the transformation properties of the ${\mf W}^0$ state, i.e.
${\bf (A57)}\,\,(1/2)mc^2=(q^0;q)$.  Furthermore by \eqref{2.29} masses are expressed
in terms of the quantum potential ${\bf (A58)}\,\,(1/2)mc^2=(p|p^0)Q^0(q^0)-Q(q)$.
In particular in \cite{f3} the role of the quantum potential was seen as a sort of
intrinsic self energy which is reminiscent of the relativistic self energy and
{\bf (A58)} provides a more explicit evidence of such an interpretation.$\hfill\bs$

\section{QUANTUM FIELD THEORY}
\renewcommand{\theequation}{3.\arabic{equation}}
\setcounter{equation}{0}

In trying to imagine particle trajectories of a fractal nature or in a
fractal medium we are tempted to abandon (or rather relax) the particle
idea and switch to quantum fields (QF).  Let the fields sense the bumps
and fractality; if one can think of fields as operator valued
distributions for example then fractal supports for example are quite reasonable. 
There are other reasons of course since the notion of particle in quantum
field theory (QFT) has a rather fuzzy nature anyway. Then of course there
are problems with QFT itself (cf.
\cite{w12}) as well as arguments that there is no first quantization
(except perhaps in the Bohm theory - cf. \cite{n57,z4}).  We review here
some aspects of particles arising from QF and QFT methods, especially in
a Bohmian spirit (cf.
\cite{b26,b3,c64,d4,d42,h97,i1,k12,m41,n61,n62,w17}).

\subsection{EMERGENCE OF PARTICLES}

We refer to \cite{h97,w12} for interesting philosophical discussion about
particles and localized objects in a QFT and will extract here from 
\cite{b26,c64,d42,n61,n62}; for QFT we refer to \cite{h92,s54}. 
We omit many details and assume standart QFT techniques are known.  First 
\cite{n62} is impressive in producing a {\bf local} operator describing the
particle density current for scalar and spinor fields in an arbitrary
gravitational and electromagnetic background.  This enables one to
describe particles in a local, general covariant, and gauge invariant
manner.  The current depends on the choice of a 2-point Wightman function
and a most natural choice based on the Green's function \`a la Schwinger-deWitt
leads to local conservation of the current provided that
interaction with quantum fields is absent.  Interactions lead to local
nonconservation of current which describes local particle production
consistent with the usual global description based on the interaction
picture. 
Thus for suitable choice of a 2-point Wightman function $W(x,x')$ the formula
for scalar fields is given by ${\bf (B1)}\,\,j_{\mu}(x)=(1/2)\int_{\gS}d\gS^{'\nu}
\{W(x,x')\olra{\pp}\!\!\!_{\mu}\olra{\pp}'\!\!\!_{\nu}\phi(x)\phi(x')+h.c.\}$.  Upon
extracting formulas for $\phi^{\pm}$ in $\phi=\phi^{+}+\phi^{-}$ in 
\bq\label{3.8}
\phi^{+}(x)=i\int_{\gS}
d\gS^{'\nu}W^{+}(x,x')\olra{\pp}'\!\!\!_{\nu}\phi(x');\,\,\phi^{-}(x)= -i\int_{\gS}
d\gS^{'\nu}W^{-}(x,x')\olra{\pp}'\!\!\!_{\nu}\phi(x')
\end{equation}
one arrives at 
\bq\label{3.9}
j_{\mu}(x)=\frac{1}{2}\int_{\gS}d\gS^{'\nu}\left[W^{+}(x,x')\olra{\pp}\!\!\!_{\mu}
\olra{\pp}'\!\!\!_{\nu}\phi(x)\phi(x;)+W^{-}(x,x')\olra{\pp}\!\!\!_{\mu}\olra{\pp}'
\!\!\!_{\nu}\phi(x')\phi(x)\right]
\end{equation}
and thence to ${\bf
(B2)}\,\,j_{\mu}=i\phi^{-}\olra{\pp}\!\!\!_ {\mu}\phi^{+}$ or ${\bf
(B3)}\,\,j_{\mu}=(i/2)N_{-}\phi\olra{\pp}\!\!\!_{\mu}\phi$ where
$N_{-}\phi^{+}\phi^{-}=-\phi^{-}\phi^{+}$.  It is also demonstrated that energy
production corresponds exactly to particle production.$\hfill\bs$

\subsection{FIELD THEORY MODELS}

\begin{example}
In the bosonic theory of \cite{n61} for a relativistic KG equation 
${\bf (B4)}\,\,(\pp_0^2-\na^2+m^2)\phi=0$. 
one has
a corresponding particle current ${\bf (B5)}\,\,j_{\mu}=i\psi^*\olra{\pp}\!\!\!_{\mu}
\psi$ and particles have a trajectory velocity ${\bf (B6)}\,\,(d\vec{x}/dt)=
\vec{j}(t,\vec{x})/j_0(t,\vec{x})$.  One obtains a HJ equation $(1/2m)\bx S-
(c^2m/2)+Q=0$ with ${\bf (B7)}\,\,Q=-(1/2m)(\pp^{\mu}\pp_{\mu}R)/R=
-(1/2m)(\bx R/R)$ (quantum potential).  In fact this leads to ${\bf
(B8)}\,\,md^2_{\tau}x_{\mu}=\pp^{\mu}Q$ and the physical number of particles $N_{phyx}
=\int d^3x|j_0|$ is not conserved (although $N=\int d^3xj_0$ is).
In an interaction picture with ${\bf
(B9)}\,\,(\pp_0^2-\na^2+m^2)\hat{\phi}=J(\hat{\phi})$ (and $c=\hbar=1$) one has a dBB
interpretation via ($\Psi\sim\Psi[\phi({\bf x}0,t)]$)
\bq\label{3.17}
(\pp_0^2-\na^2)\phi(x)=J(\phi(x))-\left(\frac{\gd Q[\phi,t]}{\gd\phi({\bf
x})}\right)_{\phi({\bf x})=\phi(x)};\,\,Q=-\frac{1}{2|\Psi|}\int d^3x\frac{\gd^2
|\Psi|}{\gd\phi^2({\bf x})}
\end{equation}
with quantum potential ${\bf (B10)}\,\,Q=-(1/2|\Psi|)
\int d^3x[\gd^2|\Psi|/\gd\phi^2(x)$ in functional form and standard physics abuse of
notation (cf. Example 3.2 to conclude that Q generates mass).   The n particles
attributed to the wave function $\psi_n$ also have trajectories given via 
\bq\label{3.18}
\frac{d{\bf x}_{n,j}}{dt}=\left(\frac{\psi^*_n(x^{(n)})\olra{\na}\!\!\!_j\psi_n
(x^{(n)})}{\psi_n^*(x^{(n)})\olra{\pp}\!\!\!_{t_j}\psi_n(x^{(n)})}
\right)_{t_1=\cdots=t_n=t}
\end{equation}
An effectivity parameter 
${\bf (B11)}\,\,e_n[\phi,t]=|\tl{\Psi}_n[\phi,t]|^2/\sum_{n'}^
{\infty}|\tl{\Psi}_{n'}[\phi,t]|^2$
is defined to be a nonlocal hidden
variable attributed to the particle introduced to provide a deterministic description
of the creation and annihilation of particles.  Here $\Psi[\phi,t]=\sum_0^{\infty}
\tl{\Psi}_n[\phi,t]$ where $\tl{\Psi}_n$ are unnormalized n-particle wave functionals
and $\psi_n({\bf x},t)=<0|\hat{\phi}(t,{\bf x})\cdots\hat{\phi}(t,{\bf x})|\Psi>$.
$\hfill\bs$
\end{example}
\begin{example}
Quantum fields are also discussed briefly in \cite{h99} and we extract here from this
source. The approach follows \cite{b70} and one takes ${\mc
L}=(1/2)\pp_{\mu}\psi\pp^{\mu}
\psi=(1/2)[\dot{\psi}^2-(\na\psi)^2]$ as Lagrangian where $\dot{\psi}=\pp_t\psi$ and 
variational technique yields the wave equation $\bx \psi=0$ ($\hbar=c=1$).  Define 
conjugate momentum as $\pi=\pp{\mc L}/\pp\dot{\psi}$, the Hamiltonian via 
${\mc H}=\pi\dot{\psi}-{\mc L}=(1/2)[\pi^2+(\na \psi))^2]$, and the field Hamiltonian by
${\mf H}=\int {\mc H}d^3x$.  Replacing $\pi$ by $\gd S/\gd\psi$ where $S[\psi]$ is a
functional the classical HJ equation of the field $\pp_tS+H=0$ becomes
\bq\label{alpha}
\frac{\pp S}{\pp t}+\frac{1}{2}\int d^3x\left[\left(\frac{\gd S}{\gd\psi}\right)^2+(\na
S)^2\right]=0
\end{equation}
The term $(1/2)\int d^3x(\na \psi)^2$ plays the role of an external potential.  To
quantize the system one treats $\psi({\bf x})$ and $\pi({\bf x})$ as Schr\"odinger
operators with $[\psi({\bf x}),\psi({\bf x}')]=[\pi({\bf x}),\pi({\bf x}')]=0$ and
$[\psi({\bf x}),\pi({\bf x}')]=i\gd({\bf x}-{\bf x}')$.  Then one works in a
representation $|\psi({\bf x})>$ in which the Hermitian operator $\psi({\bf x})$ is
diagonal.  The Hamiltonian becomes an operator $\hat{H}$ acting on a wavefunction
$\Psi[\psi({\bf x}),t)=<\psi({\bf x})|\Psi(t)>$ which is a functional of the real field
$\psi$ and a function of t.  This is not a point function of ${\bf x}$ since $\Psi$
depends on the variable $\psi$ for all ${\bf x}$.  Now the SE for the field is 
$i\pp_t\Psi=\hat{H}\Psi$ or explicitly
\bq\label{beta}
i\frac{\pp\Psi}{\pp t}=\int d^3x\frac{1}{2}\left[-\frac{\gd^2}{\gd\psi^2}+(\na \psi)^2
\right]\Psi
\end{equation}
Thus $\psi$ is playing the role of the space variable ${\bf x}$ in the particle SE and
the continuous index ${\bf x}$ here is analogous to the discrete index i in the many
paricle theory.  To arrive at a causal interpretation now one writes $\Psi=Rexp(iS)$ for
$R,S[\psi,t]$ real functionals and decomposes \eqref{beta} as
\bq\label{gamma}
\frac{\pp S}{\pp t}+\frac{1}{2}\int d^3x\left[\left(\frac{\gd S}{\gd\psi}\right)^2+(\na
\psi)^2\right]+Q=0;\,\,\frac{\pp R^2}{\pp t}+\int d^3x\frac{\gd}{\gd\psi}\left(R^2\frac
{\gd S}{\gd \psi}\right)=0
\end{equation}
where the quantum potential is now ${\bf (B12)}\,\,Q[\psi,t]=-(1/2R)\int d^3x(\gd^2R/
\gd\psi^2)$.  \eqref{gamma} now gives a conservation law wherein at time t $R^2D\psi$ is
the probability for the field to lie in an element of volume $D\psi$ around $\psi$,
where $D\psi$ means roughly $\prod_{\bf x}d\psi$ and there is a normalization
$\int|\Psi|^2D\psi=1$.  Now introduce the assumption that at each instant t the field
$\psi$ has a well defined value for all ${\bf x}$ as in classical field theory,
whatever the state $\Psi$.  Then the time evolution is obtained from the solution of
the ``guidance" formula
\bq\label{delta}
\frac{\pp\psi({\bf x},t)}{\pp t}=\left.\frac{\gd S[\psi({\bf x},t]}{\gd\psi({\bf x})}
\right|_{\psi({\bf x})=\psi({\bf x},t)}
\end{equation}
(analogous to $m\ddot{{\bf x}}=\na S$) once one has specified the initial function 
$\psi_0({\bf x})$ in the HJ formalism.  To find the equation of motion for the field
coordinates apply $\gd/\gd\psi$ to the HJ equation \eqref{gamma} to get
\bq\label{epsilon}
\frac{d}{dt}\dot{\psi}=-\frac{\gd}{\gd \psi}\left[Q+\frac{1}{2}\int d^3x(\na
\psi)^2\right];\,\,\frac{d}{dt}=\frac{\pp}{\pp t}+\int d^3x\frac{\pp\psi}{\pp t}\frac
{\gd}{\gd\psi}
\end{equation}
This is analogous to $m\ddot{{\bf x}}=-\na(V+Q)$ and, noting that
$d\dot{\psi}/dt=\pp\dot{\psi}/dt$ and taking the classical external force term to the
right one arrives at
\bq\label{mu}
\bx\psi({\bf x},t)=-\left.\frac{\gd Q[\psi({\bf x},t]}{\gd\psi({\bf x})}\right|_
{\psi({\bf x})=\psi({\bf x},t)}
\end{equation}
The quantum force term on the right side is responsible for all the characteristic
effects of QFT.  In particular comparing to a classical massive KG equation $\bx \psi
+m^2\psi=0$ with suitable initial conditions one can argue that the quantum force
generates mass in the sense that the massless quantum field acts as if it were a
classical field with mass given via the quantum potential (cf. Remark 2.1).$\hfill\bs$
\end{example}
\begin{example}
In the fermionic theory of \cite{n61} one defines trajectory velocities 
\bq\label{3.19}
\frac{d{\bf x}^P}{dt}=\frac{{\bf j}^P(t,{\bf x}^P)}{j_0^P(t,{\bf x}^P)};\,\,\frac
{d{\bf x}^A}{dt}=\frac{{\bf j}^A(t,{\bf x}^A)}{j_0^A(t,{\bf x}^A)}
\end{equation}
for a causal interpretation of the Dirac equation.  For the field theory the Grassman
fields are bosonized in terms of $\phi$ fields and it is shown how to create sources
and velocities leading to an equation ${\bf
(B13)}\,\,d\vec{\phi}/dt=\vec{v}(\vec{\phi},t)$.  This is made explicit and
effectivity parameters are defined again (cf. \cite{n61} for details and
philosophy).$\hfill\bs$
\end{example}
\begin{example}
Going now to \cite{d4,d42,d15} the philosophy revolves around Bell type QFT containing
the idea of stochastic jumps via Markov processes but expressed field theoretically. 
(We refer to \cite{n15} for the diffusion approach to relativistic QM via Markov
processes with jumps; the approach is however very different.)  We mention also again
that the KG equation is not covered (yet) in this theory (cf. however \cite{b3,h99}). 
The central formula is
\bq\label{4.6}
\gs(dq|q')=\frac{[(2/\hbar)\Im<\Psi|P(dq)HP(dq')|\Psi>]^{+}}
{<\Psi|P(dq')|\Psi>}
\end{equation}
which describes the jump probability in a space ${\mf Q}=\cup{\mf Q}^n$ of
world lines.  Transition probabilities for the Markov process $Q_t$ are described by
forward and backward generators ${\mc L}_t$ and $L_t$ which are dual via 
${\bf (B14)}\,\,\int f(q){\mc L}_t\rho(dq)=\int
L_tf(q)\rho (dq)$. 
One looks for equivariant transition probabilities so that $|\Psi_t|^2=\rho_t$ for all
$t$ corresponds to ${\mc L}_t\rho_t=\pp_t\rho_t=\pp_t|\Psi|^2$.  Jump processes will
correspond to integral operator type Hamiltonians and one will have jump relations 
${\bf (B15)}\,\,{\mc L}\rho(dq)=\int_{q'\in{\mf Q}}
(\gs(dq|q')\rho(dq')-\gs(dq')|q)\rho(dq))$
subsequent to which one goes through various constructions involving positive
operator valued measures.  We refer to \cite{d4,d42,d15} for details and many examples
(cf. also \cite{b1,b2,b5,d13,g1,v1}
for information on Bohmian theory).
$\hfill\bs$
\end{example}
\indent
{\bf REMARK 3.1.}
For a discussion of a possible quantum origin of the gravitational interaction via the
quantum potential see \cite{m7}.$\hfill\bs$
\\[3mm]\indent
{\bf REMARK 3.2.}
There are a number of papers by T. Arimitsu et al dealing with quantum stochastic
diffusion equations for boson and fermion systems in the context of nonequilibrium
thermo field dynamics (NETFD).  The materia also goes into the thermodynamics of
multifractal systems and turbulence and we refer here to \cite{a2,a3,j2,k53}.  This
material seems very interesting but we postpone discussion for now.
$\hfill\bs$
\\[3mm]\indent
{\bf REMARK 3.3.}
We mention also \cite{k1,k2,k4} where some interesting thermal equations arise related
to Schr\"odinger equations, Klein-Gordon equations, etc.  One deals with systems (micro
or macro) having a thermal history described via ${\bf (B16)}\,\,q(t)=\int_{-\infty}^t
K(t-t')\na T(t')dt'$.  Here $q(t)$ is the density of the energy flux and K describes
the thermal memory, often of the form ${\bf (B17)}\,\,K(t-t')=(K/\tau)exp[-(t-t')/
\tau]$ where K is constant and $\tau$ denotes the relaxation time.  There are three
principal situations 
\bq\label{5.1}
K(t-t')=\left\{\begin{array}{cc}
K\,lim_{t_0\to 0}\gd(t-t'-t_0) & diffusion\\
K=constant & wave\\
(K/\tau)exp\left[-\frac{(t-t')}{\tau}\right] & damped\,\,wave\,\,or\,\,hyperbolic\,\,
diffusion
\end{array}\right.
\end{equation}
The damped wave or hyperbolic diffusion equation is ${\bf
(B18)}\,\,\pp_t^2T+(1/\tau)\pp_tT=(D_T/\tau)\na^2T$ which for $\tau\to 0$ becomes
${\bf (B19)}\,\,\pp_tT=D_T \na^2T$ where $D_T$ is the thermal diffusion coefficient.
The systems with very short relaxation times have very short memory.  For
$\tau\to\infty$ {\bf (B18)} has the form of an undamped thermal wave equation or
ballistic thermal equation.  In solid state physics the ballistic phonons or electrons
are those for which $\tau\to\infty$.  Experiments with ballistic phonons or electrons
demonstrate the existence of wave motion on the lattice scale or on the electron gas
scale ${\bf (B20)}\,\,\pp_t^2T=(D_T/\tau)\na^2T$.  Now define in {\bf (B18)} the
quantity
${\bf (B21)}\,\,v=(D_T/\tau)^{1/2}$ (velocity of thermal wave propagation) and 
${\bf (B22)}\,\,\gl=v\tau$ ($\gl$ is the mean free path of the heat carriers).  Then
{\bf (B18)} can be written as ${\bf (B23)}\,\,(1/v^2)\pp_t^2/t+(1/\tau
v^2)\pp_tT=\na^2T$. Formally with substitutions $t\leftrightarrow it$ and
$T\leftrightarrow \psi$ this is
${\bf (B24)}\,\,i\hbar\psi_t=-(\hbar^2/2m)\na^2\psi-\hbar\tau\psi_{tt}$ where one uses
$D_T=\hbar^2/2m$ and $\tau=\hbar/2mv^2$.  One could also embellish this with a potential
term $V\psi\sim VT$ to get
\bq\label{5.2}
i\hbar\psi_t=-\frac{\hbar^2}{2m}\na^2\psi+V\psi-\hbar\tau\psi_{tt}
\end{equation}
The term $\tau\hbar\psi_{tt}$ could be envisioned as Zitterbewegung describing the
interaction of ``thermal particles" (say electrons) with ``spacetime" $\sim$ a vacuum
full of virtual particle pairs (say electron-positron pairs).  One can argue that in
certain realistic situations ${\bf (B25)}\,\,\tau\sim\tau_{P}=\tau_{Planck}=
(1/2)(\hbar G/c^5)^{1/2}=\hbar/2M_Pc^2$ where $M_P\sim$ Planck mass; then \eqref{5.2}
can be written as 
\bq\label{5.3}
i\hbar\frac{\pp\psi}{\pp
t}=-\frac{\hbar^2}{2m}\na^2\psi+V\psi-\frac{\hbar^2}{2M_P}\na^2\psi+\frac{\hbar^2}{2M_P}
\left(\na^2\psi-\frac{1}{c^2}\psi_{tt}\right)
\end{equation}
where the last two terms with ${\bf (B26)}\,\,\na^2\psi-(1/c^2)\psi_{tt}=0$ could
perhaps be considered as describing a Bohmian type pilot wave (cf. \cite{b40,b38,b8}
for general philosophy).  Note that the pilot wave does not depend on the paricle mass
$m$ (nor on $M_P$) and for $m<<M_P$ Schr\"odinger mechanics prevails; however the pilot
wave still exists.
$\hfill\bs$

\section{SCALE RELATIVITY}
\renewcommand{\theequation}{4.\arabic{equation}}
\setcounter{equation}{0}

In \cite{c7} and Example 2.6 here we sketched a few developments in the theory of
scale relativity.  This is by no means the whole story and we want to give a taste of
the main ideas along with deriving KG and Dirac equations in this context (cf. 
\cite{a10,c20,c13,c31,n4,n5,n55,n13,n30,n31}).  A main idea here is that the
Schr\"odinger, Klein-Gordon, and Dirac equations are all geodesic equations in the
fractal framework.  They have the form $D^2/ds^2=0$ where $D/ds$ represents the
appropriate covariant derivative.  The complex nature of the SE and KGE aris from a
discrete time symmetry breaking based on nondifferentiability.  For the Dirac
equation further discrete symmetry breakings are needed on the spacetime variables
in a biquaternionic context (cf. here \cite{c20}).  First we go back to
\cite{n5,n30,n31} and sketch some of the fundamentals of scale relativity.  This is a
very rich and beautiful theory extending in both spirit and generality the relativity
theory of Einstein but it does not yet appear to have been sanctified by the
extablishment (cf. also
\cite{c99} for variations involving Clifford theory). The basic idea here is that (following
Einstein) the laws of nature apply whatever the state of the system and hence the relevant
variables can only be defined relative to other states.  Standard scale laws of power-law
type correspond to Galilean scale laws and from them one actually recovers quantum mechanics
(QM) in a nondifferentiable space.  The quantum behavior is a manifestation of the fractal
geometry of spacetime. In particular (as indicated in Example 3.6) the quantum
potential is a manifestation of fractality in the same way as the Newton potential is
a manifestation of spacetime curvature.  In this spirit one can also conjecture (cf.
\cite{n30}) that this quantum potential may explain various dynamical effects
presently attributed to dark matter (cf. also \cite{a35}).  Now for basics one deals
with a continuous but nondifferentiable physics.  It is known for example that the
length of a continuous nondifferentiable curve is dependent on the resolution $\gep$. 
One approach now involves smoothing a nondifferentiable function $f$ via ${\bf
(C1)}\,\,f(x,\gep)=
\int_{-\infty}^{\infty}\phi(x,y,\gep)f(y)dy$ where $\phi$ is smooth and say
``centered" at $x$ (we refer also to \cite{a10,c13,c31} for a refined treatment of
such matters).  There will now arise differential equations involving $\pp f/\pp log
(\gep)$ and $\pp^2f/\pp x\pp log(\gep)$ for example and the $log(\gep)$ term arises
as follows.  Consider an infinitesimal dilatation $\gep\to\gep'=\gep(1+d\rho)$ with a
curve length ${\bf (C2)}\,\,\ell(\gep)\to\ell(\gep')=\ell(\gep+\gep d\rho)=\ell(\gep)
+\gep\ell_{\gep}d\rho=(1+\tl{D}d\rho)\ell(\gep)$.  Then ${\bf (C3)}\,\,\tl{D}=\gep
\pp_{\gep}=\pp/\pp log(\gep)$ is a dilatation operator and in the spirit of
renormalization (multiscale approach) one can assume ${\bf
(C4)}\,\,\pp\ell(x,\gep)/\pp log(\gep)=\gb(\ell)$ (where $\ell(x,\gep)$ refers to the
curve defined by $f(x,\gep)$).  Now for Galilean scale relativity consider ${\bf
(C5)}\,\,\pp\ell(x,\gep)/\pp log(\gep)=a+b\ell$ which has a solution
\bq\label{a1}
\ell(x,\gep)=\ell_0(x)\left[1+\gz(x)\left(\frac{\gl}{\gep}\right)^{-b}\right]
\end{equation}
where $\gl^{-b}\gz(x)$ is an integration constant and $\ell_0=-a/b$.  One can choose
$\gz(x)$ so that $<\gz^2(x)>=1$ and for $a\ne 0$ there are two regimes (for $b<0$)
\begin{enumerate}
\item
$\gep<<\gl\Rightarrow \gz(x)(\gl/\gep)^{-b}>>1$ and $\ell$ is given by a scale
invariant fractal like power with dimension $D=1-b$, namely $\ell(x,\gep)=\ell_0
(\gl/\gep)^{-b}$.
\item
$\gep>>\gl\Rightarrow \gz(x)(\gl/\gep)^{-b}<<1$ and $\ell$ is independent of scale.
\end{enumerate}
\indent
Here $\gep=\gl$ constitutes a transition point between fractal and nonfractal
behavior.
Only the special case $a=0$ yields unbroken scale invariance of $\ell=\ell_0
(\gl/\gep)^{\gd}\,\,(\gd=-b)$ and {\bf (C5)} becomes then ${\bf (C6)}\,\,
\tl{D}\ell=b\ell$ so the scale dimension is an eigenvalue of $\tl{D}$.  Finally the
case $b>0$ corresponds to the  cosmological domain.
\\[3mm]\indent
Now one looks for scale covariant laws and checks this for power laws ${\bf (C7)}\,\,
\phi=\phi_0(\gl/\gep)^{\gd}$.  Thus a scale transformation for $\gd(\gep')=\gd(\gep)$
will have the form
\bq\label{a2} 
log\frac{\phi(\gep')}{\phi_0}=log\frac{\phi(\gep)}{\phi_0}+V\gd(\gep);\,\,
V=log\frac{\gep}{\gep'}
\end{equation}
In the same way that only velocity differences have a physical meaning in Galilean
relativity here only V differences or scale differences have a physical meaning. 
Thus V is a ``state of scale" just as velocity is a state of motion.  In this spirit
laws of linear transformation of fields in a scale transformation $\gep\to\gep'$
amount to finding $A,B,C,D(V)$ such that
\bq\label{a3}
log\frac{\phi(\gep')}{\phi_0}=A(V)log\frac{\phi(\gep)}{\phi_0}+B(V)\gd(\gep);\,\,
\gd(\gep')=C(V)log\frac{\phi(\gep)}{\phi_0}+D(V)\gd(\gep)
\end{equation}
Here $A=1,\,B=V,\,C=0,\,D=1$ corresponds to the Galileo group.  Note also ${\bf
(C8)}\,\,\gep\to\gep'\to\gep''\Rightarrow V''=V+V'$.  For the analogue of Lorentz
transformations there is a need to preserve the Galilean dilatation law for scales
larger than the quantum classical transition.  Note $V=log(\gep/\gep')\sim
\gep/\gep'=exp(-V)$ and set $\rho=\gep'/\gep$ with $\rho'=\gep''/\gep'$ and
$\rho''=\gep''/\gep$; then ${\bf (C9)}\,\,log\rho''=\log\rho+\log\rho'$ and one is
thinking here of $\rho:\,\gep\to\gep',\,\,\rho':\,\gep'\to\gep''$ and 
$\rho'':\,\gep\to\gep''$ with compositions (the notation is meant to 
correspond to {\bf (C2)}).  We recall the Einstein-Lorentz law 
${\bf (C10)}\,\,w=(u+v)/[1+(uv/c^2)]$ but one now has several regimes to consider.
Following \cite{n31} small scale symmetry is broken by mass via the emergence of ${\bf
(C11)}\,\,\gl=
\gl_c=\hbar/mc$ (Compton length) and $\gl_{dB}=\hbar/m\nu$ (deBroglie length), while
for extended objects $\gl_{th}=\hbar/m<\nu^2>^{1/2}$ (thermal deBroglie length)
determines the transition scale.  The transition scale in \eqref{a1} is the
Einstein-deBroglie scale (in rest frame $\gl\sim\tau=\hbar/mc^2$) and in the cosmological
realm the scale symmetry is broken by the emergence of static structure of typical size
$\gl_g= GM/3<\nu^2>$.  The scale space consists of three domains (quantum, classical -
scale independent, and cosmological).  Another small scale transition factor appears
in the Planck length scale $\gl_P=(\hbar G/c^3)^{1/2}$ and at large scales the
cosmological constant $\gL$ comes into play.  With this background the composition
of dilatations is taken to be
\bq\label{a4}
log\frac{\gep'}{\gl}=\frac{log\rho+\log\frac{\gep}{\gl}}{1+\frac{log\rho
log\frac{\gep}{\gl}}{log^2(L/\gl)}}=\frac{log\rho+\log\frac{\gep}{\gl}}{1+
\frac{log\rho log(\gep/\gl)}{C^2}}
\end{equation}
where $L\sim \gl_P$ near small scales and $L\sim\gL$ near large scales
(note $\gep=L\Rightarrow \gep'=L$ in \eqref{a4}).  Comparing
with {\bf (C10)} one thinks of $log(L/\gl)=C\sim c$ (note here
$log^2(a/b)=log^2(b/a)$ in comparing formulas in \cite{n30,n31}).  
Lengths now change via 
\bq\label{a5}
log\frac{\ell'}{\ell_0}=\frac{log(\ell/\ell_0)+\gd
log\rho}{\sqrt{1-\frac{\log^2\rho}{C^2}}}
\end{equation}
and the scale variable $\gd$ (or djinn) is no longer constant but changes via
\bq\label{a6}
\gd(\gep')=\frac{\gd(\gep)+\frac{log\rho log(\ell/\ell_0)}{C^2}}{\sqrt{1-\frac
{log^2\rho}{C^2}}}
\end{equation}
where $\gl\sim$ fractal-nonfractal transition scale.  
\\[3mm]\indent
We have derived the SE in Example 2.6 (cf. also \cite{c7}) and go now to the KG
equation via scale relativity.  The derivation in the first paper of \cite{c20}
seems the most concise and we follow that (cf. also \cite{n31}).  All of the
elements of the approach for the SE remain valid in the motion relativistic case with
the time replaced by the proper time s, as the curvilinear parameter along the 
geodesics.  Consider a small increment $dX^{\mu}$ of a nondifferentiable four
coordinate along one of the geodesics of the fractal spacetime.  One can decompose
this in terms of a large scale part $\overline{LS}<dX^{\mu}>=dx^{\mu}=v_{\mu}ds$ and
a fluctuation $d\xi^{\mu}$ such that $\overline{LS}<d\xi^{\mu}>=0$.  One is led to
write the displacement along a geodesic of fractal dimension $D=2$ via
${\bf (C12)}\,\,dX^{\mu}_{\pm}=d_{\pm}x^{\mu}+d\xi^{\mu}_{\pm}=v_{\pm}^{\mu}ds+
u_{\pm}^{\mu}\sqrt{2{\mc D}}ds^{1/2}$.  Here $u_{\pm}^{\mu}$ is a dimensionless
fluctuation andd the length scale $2{\mc D}$ is introduced for dimensional purposes.
The large scale forward and backward derivatives $d/ds_{+}$ and $d/ds_{-}$ are
defined via 
\bq\label{a7}
\frac{d}{ds_{\pm}}f(s)=lim_{s\to 0_{\pm}}\ol{LS}\left<\frac{f(s+\gd s)-f(s)}
{\gd s}\right>
\end{equation}
Applied to $x^{\mu}$ one obtains the forward and backward large scale four velocities
of the form
${\bf (C13)}\,\,(d/dx_{+})x^{\mu}(s)=v_{+}$ and $(d/ds_{-})x^{\mu}=v_{-}^{\mu}$.
Combining yields 
\bq\label{a8}
\frac{d'}{ds}=\frac{1}{2}\left(\frac{d}{ds_{+}}+\frac{d}{ds_{-}}\right)-\frac{i}{2}
\left(\frac{d}{ds_{+}}-\frac{d}{ds_{-}}\right);
\end{equation}
$${\mc V}^{\mu}=\frac{d'}{ds}x^{\mu}=
V^{\mu}-iU^{\mu}=\frac{v_{+}^{\mu}+v_{-}^{\mu}}{2}-i\frac{v_{+}^{\mu}-v_{-}^{\mu}}{2}$$
For the fluctuations one has ${\bf
(C14)}\,\,\ol{LS}<d\xi_{\pm}^{\mu}d\xi_{\pm}^{\nu}>=\mp 2{\mc D}\eta^{\mu\nu}ds$.
One chooses here $(+,-,-,-)$ for the Minkowski signature for $\eta^{\mu\nu}$ and
there is a mild problem because the diffusion (Wiener) process makes sense only for
positive definite metrics.  Various solutions were given in \cite{d97,s99,z13} and
they are all basically equivalent, amounting to the transformatin a Laplacian into a 
D'Alembertian.  Thus the two forward and backward differentials of $f(x,s)$ should be
written as ${\bf (C15)}\,\,(df/ds_{\pm})=(\pp_s+v_{\pm}^{\mu}\pp_{\mu}\mp{\mc D}
\pp^{\mu}\pp_{\mu})f$.  One considers now only stationary functions f, not depending
explicitly on the proper time s, so that the complex covariant derivative operator
reduces to ${\bf (C16)}\,\,(d'/ds)=({\mc V}^{\mu}+i{\mc D}\pp^{\mu})\pp_{\mu}$.
\\[3mm]\indent
Now assume that the large scale part of any mechanical system can be characterized by
a complex action ${\mf S}$ leading one to write ${\bf (C17)}\,\,\gd{\mf
S}=-mc\gd\int_a^bds=0$ where $ds=\ol{LS}<\sqrt{dX^{\nu}dX_{\nu}}>$.  This leads to
${\bf (C18)}\,\,\gd{\mf S}=-mc\int_a^b{\mc V}_{\nu}d(\gd x^{\nu})$ with $\gd x^{\nu}
=\ol{LS}<dX^{\nu}>$.  Integrating by parts yields ${\bf (C19)}\,\,\gd{\mf S}=-[mc\gd
x^{\nu}]_a^b+mc\int_a^b\gd x^{\nu}(d{\mc V}_{\mu}/ds)ds$.  To get the equations of
motion one has to determine $\gd {\mf S}=0$ between the same two points, i.e.
at the limits $(\gd x^{\nu})_a=(\gd x^{\nu})_b=0$.  From {\bf (C19)} one obtains then
a differential geodesic equation ${\bf (C20)}\,\,d{\mc V}/ds=0$.  One can also write
the elementary variation of the action as a functional of the coordinates.  So
consider the point a as fixed so $(\gd x^{\nu})_a=0$ and consider b as variable.
The only admissable solutions are those satisfying the equations of motion so the
integral in {\bf (C19)} vanishes and writing $(\gd x^{\nu})_b$ as $\gd x^{\nu}$ gives
${\bf (C21)}\,\,\gd{\mf S}=-mc{\mc V}_{\nu}\gd x^{\nu}$ (the minus sign comes from
the choice of signature).  The complex momentum is now ${\bf (C22)}\,\,{\mc P}_{\nu}=
mc{\mc V}_{\nu}=-\pp_{\nu}{\mf S}$ and the complex action completely characterizes the
dynamical state of the particle.  Hence introduce a wave function $\psi=exp(i{\mf S}/
{\mf S}_0)$ and via {\bf (C22)} one gets ${\bf (C23)}\,\,{\mc V}_{\nu}=(i{\mf
S}_0/mc)\pp_{\nu}log(\psi)$.  Now for the scale relativistic prescription replace the
derivative in {\bf (C20)} by its covariant expression {\bf (C16)}.  Using {\bf (C23)}
one transforms {\bf (C20)} into 
\bq\label{a9}
-\frac{{\mf S}_0^2}{m^2c^2}\pp^{\mu}log(\psi)\pp_{\mu}\pp_{\nu}log(\psi)-
\frac{{\mf S}_0{\mc D}}{mc}\pp^{\mu}\pp_{\mu}\pp_{\nu}log(\psi)=0
\end{equation}
The choice ${\mf S}_0=\hbar=2mc{\mc D}$ allows a simplification of \eqref{a9} when
one uses the identity
\bq\label{a10}
\frac{1}{2}\left(\frac{\pp_{\mu}\pp^{\mu}\psi}{\psi}\right)=\left(\pp_{\mu}log(\psi)
+\frac{1}{2}\pp_{\mu}\right)\pp^{\mu}\pp^{\nu}log(\psi)
\end{equation}
Dividing by ${\mc D}^2$ one obtains the equation of motion for the free particle 
${\bf (C24)}\,\,\pp^{\nu}[\pp^{\mu}\pp_{\mu}\psi/\psi]=0$.  Therefore the KG
equation (no electromagnetic field) is ${\bf
(C25)}\,\,\pp^{\mu}\pp_{\mu}\psi+(m^2c^2/\hbar^2)\psi=0$ and this becomes an integral
of motion of the free particle provided the integration constant is chosen in
terms of a squared mass term $m^2c^2/\hbar^2$.  Thus the quantum behavior described
by this equation and the probabilistic interpretation given to $\psi$ is reduced
here to the description of a free fall in a fractal spacetime, in analogy with
Einstein's general relativity.  Moreover these equations are covariant since the
relativistic quantum equation written in terms of $d'/ds$ has the same form as the
equation of a relativistic macroscopic and free particle using $d/ds$.  One notes
that the metric form of relativity, namely $V^{\mu}V_{\mu}=1$ is not conserved in
QM and it is shown in \cite{p5} that the free particle KG equation expressed in
terms of ${\mc V}$ leads to a new equality ${\bf (C26)}\,\,{\mc V}^{\mu}{\mc
V}_{\mu}+2i{\mc D}\pp^{\mu}{\mc V}_{\mu}=1$.  In the scale relativistic framework
this expression defines the metric that is induced by the internal scale structures
of the fractal spacetime.  In the absence of an electromagnetic field ${\mc
V}^{\mu}$ and ${\mf S}$ are related by {\bf (C22)} which can be writen as ${\bf
(C27)}\,\,{\mc V}_{\mu}=-(1/mc)\pp_{\mu}{\mf S}$ so {\bf (C26)} becomes ${\bf
(C28)}\,\,\pp^{\mu}{\mf S}\pp_{\mu}{\mf S}-2imc{\mc D}\pp^{\mu}\pp_{\mu}{\mf
S}=m^2c^2$ which is the new form taken by the Hamilton-Jacobi equation.
\\[3mm]\indent
{\bf REMARK 4.1.}
We go back to \cite{n31,p5} now and repeat some of their steps in a perhaps more
primitive but revealing form.  Thus one omits the $\ol{LS}$ notation and uses
$\gl\sim 2{\mc D}$; equations {\bf (C12)}-{\bf (C16)} and \eqref{a8} are the same
and one writes now ${\mf d}/ds$ for $d'/ds$.  Then ${\mf d}/ds={\mc
V}^{\mu}\pp_{\mu}+(i\gl/2)\pp^{\mu}\pp_{\mu}$ plays the role of a scale covariant
derivative and one simply takes the equation of motion of a free relativistic
quantum particle to be given as ${\bf (C29)}\,\,({\mf d}/ds){\mc V}^{\nu}=0$, which
can be interpreted as the equations of free motion in a fractal spacetime or as
geodesic equations.  In fact now {\bf (C29)} leads directly to the KG equation
upon writing $\psi=exp(i{\mf S}/mc\gl)$ and ${\mf P}^{\mu}=-\pp^{\mu}{\mf S}=
mc{\mc V}^{\mu}$ so that $i{\mf S}=mc\gl log(\psi)$ and ${\mc V}^{\mu}=
i\gl\pp^{\mu}log(\psi)$.  Then
\bq\label{a11}
\left({\mc
V}^{\mu}\pp_{\mu}+\frac{i\gl}{2}\pp^{\mu}\pp_{\mu}\right)\pp^{\nu}log(\psi)=0=
i\gl\left(\frac{\pp^{\mu}\psi}{\psi}\pp_{\mu}+
\frac{1}{2}\pp^{\mu}\pp_{\mu}\right)\pp^{\nu}\log(\psi)
\end{equation}
Now some identities are given in \cite{p5} for aid in calculation here, namely
\bq\label{a12}
\frac{\pp^{\mu}\psi}{\psi}\pp_{\mu}\frac{\pp^{\nu}\psi}{\psi}=
\frac{\pp^{\mu}\psi}{\psi}\pp^{\nu}\left(\frac{\pp_{\mu}\psi}{\psi}\right)=
\end{equation} 
$$=\frac{1}{2}\pp^{\nu}\left(\frac{\pp^{\mu}\psi}{\psi}\frac{\pp_{\mu}\psi}{\psi}
\right);\,\,\pp_{\mu}
\left(\frac{\pp^{\mu}\psi}{\psi}\right)+\frac{\pp^{\mu}\psi}{\psi}\frac{\pp_{\mu}\psi}
{\psi}=\frac{\pp^{\mu}\pp_{\mu}\psi}{\psi}$$
The first term in the last equation of \eqref{a11} is then
$(1/2)[(\pp^{\mu}\psi/
\psi)(\pp_{\mu}\psi/\psi)]$ and the second is
\bq\label{a13}
(1/2)\pp^{\mu}\pp_{\mu}\pp^{\nu}log(\psi)=(1/2)\pp^{\mu}\pp^{\nu}\pp_{\mu}log(\psi)=
\end{equation}
$$=(1/2)\pp^{\nu}\pp^{\mu}\pp_{\mu}log(\psi)=(1/2)\pp^{\nu}
\left(\frac{\pp^{\mu}\pp_{\mu}\psi}{\psi}-\frac{\pp^{\mu}\psi\pp_{\mu}\psi}{\psi^2}
\right)$$
Combining we get ${\bf (C30)}\,\,(1/2)\pp^{\nu}(\pp^{\mu}\pp_{\mu}\psi/\psi)=0$
which integrates then to a KG equation ${\bf
(C31)}\,\,-(\hbar^2/m^2c^2)\pp^{\mu}\pp_{\mu}\psi=\psi$ for suitable choice of
integration constant (note $\hbar/mc$ is the Compton wave length).
\\[3mm]\indent
Now in this context or above we refer back to Section 2.2 and write $Q=-(1/2m)
(\bx R/R)$ (cf. {\bf (A50)} and take $\hbar=c=1$ for convenience here).  Then recall
${\bf (C32)}\,\,\psi=exp(i{\mf S}/m\gl)$ and ${\mf P}_{\mu}=m{\mc
V}_{\mu}=-\pp_{\mu}{\mf S}$ with $i{\mf S} =m\gl log(\psi)$.  Also ${\bf (C33)}\,\,
{\mc V}_{\mu}=-(1/m)\pp_{\mu}{\mf S}=i\gl\pp_{\mu}log(\psi)$ with $\psi=Rexp
(iS/m\gl)$ so $log(\psi)=i{\mf S}/m\gl=log(R) +iS/m\gl$, leading to 
\bq\label{a14}
{\mc V}_{\mu}=
i\gl[\pp_{\mu}log(R)+(i/m\gl)\pp_{\mu}S]=-\frac{1}{m}\pp_{\mu}S+i\gl\pp_{\mu}log(R)
=V_{\mu}+iU_{\mu}
\end{equation}
Then $\bx=\pp^{\mu}\pp_{\mu}$ and $U_{\mu}=\gl\pp_{\mu}log(R)$ leads to ${\bf
(C34)}\,\,\pp^{\mu}U_{\mu}=\gl\pp^{\mu}\pp_{\mu}log(R)=\gl\bx log(R)$.
Further $\pp^{\mu}\pp_{\nu}log(R)=(\pp^{\mu}\pp_{\nu}R/R)-(R_{\nu}R_{\mu}/R^2)$ so
${\bf (C35)}\,\,\bx log(R)=\pp^{\mu}\pp_{\mu}log(R)=(\bx R/R)-(\sum
R_{\mu}^2/R^2)=(\bx R/R)-\sum (\pp_{\mu}R/R)^2=(\bx R/R)-|U|^2$ for $|U|^2=
\sum U_{\mu}^2$.  Hence via $\gl=1/2m$ for example one has 
\bq\label{a15}
Q=-(1/2m)(\bx R/R)=-\frac{1}{2m}\left[
|U|^2+\frac{1}{\gl}\bx log(R)\right]=
\end{equation}
$$=-\frac{1}{2m}\left[|U|^2+\frac{1}{\gl}\pp^{\mu}
U_{\mu}\right]=-\frac{1}{2m}|U|^2-\frac{1}{2}div(\vec{U})$$
(cf. Example 2.6 and Section 2.2).
$\hfill\bs$
\\[3mm]\indent
{\bf REMARK 4.2.}
The words fractal spacetime as used in the scale relativity methods of Nottalle et al
for producing geodesic equations (SE or KG equation) are somewhat misleading in that
essentially one is only looking at continuous nondifferentiable paths for example.
Scaling as such is of course considered extensively at other times as partially
indicated in Section 4.  It would be nice to create a fractal derivative based on
scaling properties and H-dimension alone for example which would permit the powerful
techniques of calculus to be used in a fractal context.  There has been of course
some work in this direction already in e.g. \cite{c91,g3,g5,h10,k23,k25,n37,p3,r1}
and we will sketch some of this below.
$\hfill\bs$

\section{FRACTAL CALCULUS}
\renewcommand{\theequation}{5.\arabic{equation}}
\setcounter{equation}{0}

We sketch first (in summary form) from \cite{p3} where a calculus based on fractal
subsets of the real line is formulated.  A local calculus based on renormalizing
fractional derivatives \`a la \cite{k23} is subsumed and embellished.  Consider first the
concept of content or $\ga$-mass for a (generally fractal) subset $F\subset
[a,b]$ (in what follows $0<\ga\leq 1$).  Then define the flag function for a set F
and a closed interval I as ${\bf (D1)}\,\,\gt(F,I)=1\,\,(F\cap I \ne \emptyset$ and
otherwise $\gt=0$.  Then a subdivision $P_{[a,b]}\sim P$ of $[a,b]\,\,(a<b)$ is a
finite set of points $\{a=x_0,x_1,\cdots,x_n=b\}$ with $x_i<x_{i+1}$.  If Q is any
subdivision with $P\subset Q$ it is called a refinement and if $a=b$ the set $\{a\}$
is the only subdivision.  Define then
\bq\label{51}
\gs^{\ga}[F,p]=\sum_0^{n-1}\frac{(x_{i+1}-x_i)^{\ga}}{\gG(\ga+1)}\gt(F,[x_i,x_{i+1}])
\end{equation}
For $a=b$ one defines $\gs^{\ga}[F,P]=0$.  Next given $\gd>0$ and $a\leq b$ the
coarse grained mass $\gag_{\gd}^{\ga}(F,a,b)$ of $F\cap [a,b]$ is given via
\bq\label{52}
\gag_{\gd}^{\ga}(F,a,b)=inf_{|P|\leq\gd}\gs^{\ga}[F,P]\,\,\,(|P|=max_{0\leq i\leq n-1}
(x_{i+1}-x_i))
\end{equation}
where the infimum is over P such that $|P|\leq\gd$.  Various more or less
straightforward  properties are:
\begin{itemize}
\item
For $a\leq b$ and $\gd_1< \gd_2$ one has
${\bf (D2)}\,\,\gag_{\gd_1}^{\ga}(F,a,b)\geq\gag_{\gd_2}^{\ga}(F,a,b)$.
\item
For $\gd>0$ and $a<b<c$ one has
${\bf (D3)}\,\,\gag_{\gd}^{\ga}(F,a,b)\leq\gag_{\gd}^{\ga}(F,a,c)$ and
$\gag_{\gd}^{\ga}(F,b,c)\leq\gag_{\gd}^{\ga}(F,a,c)$.
\item
${\bf (D4)}\,\,\gag_{\gd}^{\ga}$ is continuous in b and a.
\end{itemize}
\indent
Now define the mass function $\gag^{\ga}(F,a,b)$ via ${\bf (D5)}\,\,\gag^{\ga}(F,a,b)=
lim_{\gd\to 0}\gag_{\gd}^{\ga}(F,a,b)$.  The following results are proved
\begin{enumerate}
\item
{\bf (D6)} If $F\cap(a,b)=\emp$ then $\gag^{\ga}(F,a,b)=0$.
\item
Let $a<b<c$ and $\gag^{\ga}(F,a,c)<\infty$.  Then ${\bf
(D7)}\,\,\gag^{\ga}(F,a,c)=\gag^{\ga}(F,a,b)+\gag^{\ga}(F,b,c)$.  Hence $\gag^{\ga}
(F,a,b)$ is increasing in b and decreasing in a.
\item
{\bf (D8)} Let $a<b$ and $\gag^{\ga}(F,a,b)\ne 0$ be finite.  If
$0<y<\gag^{\ga}(F,a,b)$ then there exists $c,\,\,a<c<b$ such that
$\gag^{\ga}(F,a,c)=y$.
Further if $\gag^{\ga}(F,a,b)$ is finite then {\bf (D9)} $\gag^{\ga}(F,a,x)$ is
continuous for $x\in (a,b)$.
\item
For $F\subset {\bf R}$ and $\gl\in{\bf R}$ let ${\bf (D9)}\,\,F+\gl=\{x+\gl;\,x\in
F\}$.  Then ${\bf (D10)}\,\,\gag^{\ga}(F+\gl,a+\gl,b+\gl)=\gag^{\ga}(F,a,b)$ and 
$\gag^{\ga}(\gl F,\gl a,\gl b)=\gl^{\ga}\gag^{\ga}(F,a,b)$.
\end{enumerate}
\indent
Now for $a_0$ an arbitrary fixed real number one defines the integral staircase
function of order $\ga$ for F is
\bq\label{53}
S_F^{\ga}(x)=\left\{\begin{array}{cc}
\gag^{\ga}(F,a_0,x) & x\geq a_0\\
-\gag^{\ga}(F,x,a_0) & otherwise
\end{array}\right.
\end{equation}
The following properties of $S_F$ are restatements of properties for $\gag^{\ga}$.
thus
\begin{itemize}
\item
$S_F^{\ga}(x)$ is increasing in $x$.
\item
If $F\cap (x,y)=0$ then $S_F^{\ga}$ is constant in $[x,y]$.
\item
$S_F^{\ga}(y)-S_F^{\ga}(x)=\gag^{\ga}(F,x,y)$.
\item
$S_F^{\ga}$ is continuous on $(a,b)$.
\end{itemize}
\indent
Now one considers the sets F for which the mass function $\gag^{\ga}(F,a,b)$ gives
the most useful information.  Indeed one can use the mass function to define a
fractal dimension.  If $0<\ga<\gb\leq 1$ one writes
\bq\label{54}
\gs^{\gb}[F,P]\leq |P|^{\gb-\ga}\gs^{\ga}[F,P]\frac{\gG(\ga+1)}{\gG(\gb+1)};\,\,
\gag_{\gd}^{\gb}(F,a,b)\leq \gd^{\gb-\ga}\gag_{\gd}^{\ga}(F,a,b)\frac{\gG(\ga+1)}
{\gG(\gb+1)}
\end{equation}
Thus in the limit $\gd\to 0$ one gets $\gag^{\gb}(F,a,b)=0$ provided
$\gag^{\ga}(F,a,b)<\infty$ and $\ga<\gb$.  It follows that $\gag^{\ga}(F,a,b)$ is
infinite up to a certain value $\ga_0$ and then jumps down to zero for $\ga>\ga_0$
(if $\ga_0<1$).  This number is called the $\gag$-dimension of F;
$\gag^{\ga_0}(F,a,b)$ may itself be zero, finite, or infinite.  To make the
definition precise one says that the $\gag$-dimension of $F\cap [a,b]$, denoted by
$dim_{\gag}(F\cap[a,b])$, is 
\bq\label{55}
dim_{\gag}(F\cap [a,b])=\left\{\begin{array}{c}
inf\{\ga;\,\,\gag^{\ga}(F,a,b)=0\}\\
sup\{\ga;\,\,\gag^{\ga}(F,a,b)=\infty\}
\end{array}\right.
\end{equation}
One shows that ${\bf (D11)}\,\,dim_H(F\cap [a,b])\leq dim_{\gag}(F\cap [a,b])$
where $dim_H$ denotes Hausdorff dimension.  Further ${\bf (D12)}\,\,dim_{\gag}(F\cap
[a,b])\leq dim_B(F\cap [a,b])$ where $dim_B$ is the box dimension.  Some further
analysis shows that {\bf (D14)} For $F\subset {\bf R}$ compact $dim_{\gag}F=dim_HF$.
\\[3mm]\indent
Next one notes that the correspondence $F\to S_F^{\ga}$ is many to one (examples from
Cantor sets) and one calls the sets giving rise to the same staircase function
``staircasewise congruent".  The equivalence class of congruent sets containing F is
denoted by ${\mc E}_F$; thus if $G\in{\mc E}_F$ it follows that $S_G^{\ga}=S_F^{\ga}$
and ${\mc E}_G^{\ga}={\mc E}_F^{\ga}$.  One says that a point x is a point of change
of f if f is not constant over any open interval $(c,d)$ containing x.  The set of
all points of change of f is denoted by $Sch(f)$.  In particular if $G\in{\mc
E}_F^{\ga}$ then $S_G^{\ga}(x)=S_F^{\ga}(x)$ so $Sch(S_G^{\ga})=Sch(S_F^{\ga})$.  
Thus if $F\subset{\bf R}$ is such that $S_F^{\ga}(x)$ is finite for all x
($\ga=dim_{\gag}F$) then $H=Sch(S_F^{\ga})\in {\mc E}_F^{\ga}$. This takes some
proving which we omit (cf. \cite{p3}).  As a consequence let $F\subset {\bf R}$ be
such that $S_F^{\ga}(x)$ is finite for all $x\in{\bf R}$ ($\ga=dim_{\gag}F$).
Then the set $H=Sch(S_F^{\ga})$ is perfect (i.e. H is closed and every point is a
limit point).  Hence given $S_F^{\ga}(x)$ finite for all x ($\ga=dim_{\gag}F$)
one calls $Sch(S_F^{\ga})$ the $\ga$-perfect representative of ${\mc E}_F^{\ga}$ and
one proves that it is the minimal closed set in ${\mc E}_F^{\ga}$.  Indeed an
$\ga$-perfect set in ${\mc E}_F^{\ga}$ is the intersection of all closed sets G in
${\mc E}_F^{\ga}$.  One can also say that if $F\subset{\bf R}$ is $\ga$-perfect and
$x\in F$ then for $y<x<z$ either $S_F^{\ga}(y)<S_F^{\ga}(x)$ or $S_F^{\ga}(x)<
S_F^{\ga}(z)$ (or both).  Thus for an $\ga$-perfect set it is assured that the values
of $S_F^{\ga}(y)$ must be different from $S_F^{\ga}(x)$ at all points y on at least
one side of x.  As an example one shows that the middle third Cantor set $C=E_{1/3}$
is $\ga$-perfect for $\ga=log(2)/log(3)=d_H(C)$ so $C=Sch(S_C^{\ga})$.
\\[3mm]\indent
Now look at F with the induced topology from ${\bf R}$ and consider the idea of 
F-continuity.
\begin{definition}
Let $F\subset{\bf R}$ and $f:\,{\bf R}\to {\bf R}$ with $x\in F$.  A number $\ell$ is
said to be the limit of f through the points of F, or simply F-limit, as $y\to x$ if
given $\gep>0$ there exists $\gd>0$ such that $y\in F$ and $|y-x|<\gd\Rightarrow
|f(y)-\ell|<\gep$.  In such a case one writes $\ell=F-limit_{y\to x}f(y)$.  A
function $f$ is F-continuous at $x\in F$ if $f(x)=F-limit_{y\to x}f(y)$ and
uniformly F-continuuous on $E\subset F$ if for $\gep>0$ there exists $\gd>0$ such that
$x\in F,\,y\in E$ and $|y-x|<\gd\Rightarrow |f(y)-f(x)|<\gep$.  One sees that if f is
F-continuous on a compact set $E\subset F$ then it is uniformly F-continuous on E.
\end{definition}
\begin{definition}
The class of functions $f:\,{\bf R}\to{\bf R}$ which are bounded on F is denoted by
$B(F)$.  Define for $f\in B(F)$ and I a closed interval
\bq\label{56}
M[f,F,I]=\left\{\begin{array}{cc}
sup_{x\in F\cap I}f(x) & F\cap I\ne\emp\\
0 & otherwise
\end{array}\right.
\end{equation}
$$m[f,F,I]=\left\{\begin{array}{cc}
inf_{x\in F\cap I}f(x) & F\cap I\ne \emp\\
0 & otherwise
\end{array}\right.$$
\end{definition}
\begin{definition}
Let $S_F^{\ga}(x)$ be finite for $x\in[a,b]$ and P be a subdivision with points
$x_0,\cdots,x_n$.  The upper $F^{\ga}$ and lower $F^{\ga}$ sums over P are given 
respectively by
\bq\label{57}
U^{\ga}[f,F,P]=\sum_0^{n-1}M[f,F,[x_i,x_{i+1}]](S^{\ga}_F(x_{i+1})-S^{\ga}_F(x_i));
\end{equation}
$$L^{\ga}[f,F,P]=\sum_0^{n-1}m[f,F,[x_i,x_{i+1}]](S_F^{\ga}(x_{i+1})-S_F^{\ga}(x_i))$$
This is sort of like Riemann-Stieltjes integration and in fact one shows that
if Q is a refinement of P then $U^{\ga}[f,F,Q]\leq U^{\ga}[f,F,P]$ and $L^{\ga}
[f,F,Q]\geq L^{\ga}[f,F,P]$.  Further $U^{\ga}[f,F,P]\geq L^{\ga}[f,F,Q]$ for any
subdivisions of $[a,b]$ and this leads to the idea of F-integrability.  Thus
assume $S_F^{\ga}$ is finite on $[a,b]$ and for $f\in B(F)$ one defines lower and
upper $F^{\ga}$-integrals via
\bq\label{58}
\ul{\int_a^b}f(x)d_F^{\ga}x=sup_PL^{\ga}[f,F,P];\,\,\ol{\int_a^b}f(x)d_F^{\ga}x=
inf_PU^{\ga}[f,F,P]
\end{equation}
One then says that $f$ is $F^{\ga}$-integrable if ${\bf (D15)}\,\,\ul{\int_a^b}
f(x)d_F^{\ga}x=\ol{\int_a^b}f(x)d_F^{\ga}x =\int_a^bf(x)d_F^{\ga}x$.
\end{definition}
\indent
One shows then
\begin{enumerate}
\item
$f\in B(F)$ is $F^{\ga}$-integrable on $[a,b]$ if and only if for any $\gep>0$ there is a
subdivision P of $[a,b]$ such that $U^{\ga}[f,F,P]<L^{\ga}[f,F,P]+\gep$.
\item
Let $F\cap [a,b]$ be compact with $S^{\ga}_F$ finite on $[a,b]$.  Let $f\in B(F)$ and 
$a<b$; then if f is F-continuous on $F\cap [a,b]$ it follows that $f$ is $F^{\ga}$-integrable
on $[a,b]$.
\item
Let $a<b$ and $f$ be $F^{\ga}$-integrable on $[a,b]$ with $c\in (a,b)$.  Then $f$ is
$F^{\ga}$-integrable on $[a,c]$ and $[c,b]$ with ${\bf
(D16)}\,\,\int_a^bf(x)d_F^{\ga}x=\int_a^cf(x)d_F^{\ga}x +\int_c^bf(x)d_F^{\ga}x$.
\item
If $f$ is $F^{\ga}$-integrable then ${\bf (D17)}\,\,\int_a^b\gl f(x)d_F^{\ga}x=\gl\int_a^b
f(x)d_F^{\ga}x$ and, for $g$ also $F^{\ga}$-integrable, ${\bf (D18)}\,\,\int_a^b(f(x)+g(x))
d_F^{\ga}x=\int_a^bf(x)d_F^{\ga}x+\int_a^bg(x)d_F^{\ga}x$.
\item
If $f,g$ are $F^{\ga}$-integrable and $f(x)\geq g(x)$ for $x\in F\cap [a,b]$ then
${\bf (D19)}\,\,\int_a^bf(x)d_F^{\ga}x\geq \int_a^bg(x)d_F^{\ga}x$.
\end{enumerate}
\indent
One specifies also ${\bf (D20)}\,\,\int_b^af(x)d_F^{\ga}x=-\int_a^bf(x)d_F^{\ga}x$ and
it is easily shown that if $\chi_F(x)$ is the characteristic function of F then
${\bf (D21)}\,\,\int_a^b\chi_F(x)d_F^{\ga}x=S_F^{\ga}(b)-S_F^{\ga}(a)$.
Now for differentiation one writes
\bq\label{59}
{\mc D}_F^{\ga}f(x)=\left\{\begin{array}{cc}
F-lim_{y\to x}\frac{f(y)-f(x)}{S_F^{\ga}(y)-S_F^{\ga}(x)} & x\in F\\
0 & otherwise
\end{array}\right.
\end{equation}
if the limit exists.  One shows then
\begin{enumerate}
\item
If ${\mc D}_F^{\ga}f(x)$ exists for all $x\in (a,b)$ then $f(x)$ is F-continuous in $(a,b)$.
\item
With obvious hypotheses ${\mc D}_F^{\ga}(\gl f(x))=\gl{\mc D}^{\ga}_Ff(x)$ and 
${\mf D}_F^{\ga}(f+g)(x)={\mc D}_F^{\ga}f(x)+{\mc D}_F^{\ga}g(x)$.  Further if
$f$ is constant then ${\mc D}_F^{\ga}f=0$.
\item
${\mf D}_F^{\ga}(S_F^{\ga}(x))=\chi_F(x)$.
\item
(Rolle's theorem) Let $f:\,{\bf R}\to {\bf R}$ be continuous with $Sch(f)\subset F$ where
F is $\ga$-perfect and assume ${\mc D}_F^{\ga}f(x)$ is defined for all $x\in [a,b]$ with
$f(a)=f(b)=0$.  Then there is a point $c\in F\cap [a,b]$ such that ${\mc D}_F^{\ga}f(c)\geq
0$ and a point $d\in F\cap [a,b]$ where ${\mc D}_F^{\ga}f(d)\leq 0$.
\end{enumerate}
\begin{example}
This is the best that can be done with Rolle's theorem since for C the Cantor set $E_{1/3}$
take $f(x)=S_C^{\ga}(x)$ for $0\leq x\leq 1/2$ and $f(x)=1-S_C^{\ga}(x)$ for $1/2<x\leq 1$.
This function is continuous with $f(0)=f(1)=0$ and the set of change ($Sch(f)$) is C.  The 
$C^{\ga}$-derivative is given by ${\mc D}_C^{\ga}f(x)=\chi_C(x)$ for $0\leq x\leq 1/2$ and
by $-\chi_C(x)$ for $1/2<x\leq 1$.  Thus $x\in C\Rightarrow {\mc D}_C^{\ga}f(x)=\pm 1\ne 0$.
\end{example}
As a corollary one has the following result: Let $f$ be continuous with
$Sch(f)\subset F$ where F is $\ga$-perfect; assume ${\mc D}^{\ga}_Ff(s)$ exists at all
points of $[a,b]$ and that $S_F^{\ga}(b)\ne S_F^{\ga}(a)$.  Then there are points $c,d\in F$
such that
\bq\label{510}
{\mc D}_F^{\ga}f(c)\geq\frac{f(b)-f(a)}{S_F^{\ga}(b)-S_F^{\ga}(b)};\,\,{\mc D}_F^{\ga}f(d)
\leq\frac{f(b)-f(a)}{S_F^{\ga}(b)-S_F^{\ga}(a)}
\end{equation}
Similarly if $f$ is continuous with $Sch(f)\subset F$ and ${\mc D}_F^{\ga}f(x)=0\,\,\forall
x\in [a,b]$ then $f(x)$ is constant on $[a,b]$.  There are also other fundamental theorems
as follows
\begin{enumerate}
\item
(Leibniz rule)  If $u,v:\,{\bf R}\to {\bf R}$ are $F^{\ga}$-differentiable then 
${\bf (D22)}\,\,{\mc D}_F^{\ga}(uv)(x)=({\mc D}_F^{\ga}u(x))v(x)+u(x){\mc D}_F^{\ga}v(x)$.
\item
Let $F\subset{\bf R}$ be $\ga$-perfect. If $f\in B(F)$ is F-continuous on
$F\cap [a,b]$ with $g(x)=\int_a^xf(y)d_F^{\ga}y$ for all $x\in [a,b]$ then
${\mc D}_F^{\ga}g(x)=f(x)\chi_F(x)$.
\item
Let $f:\,{\bf R}\to{\bf R}$ be continuous and $F^{\ga}$-differentiable with $Sch(f)$
contained in an $\ga$-perfect set F; let also $h:\,{\bf R}\to{\bf R}$ be F-continuous
such that $h(x)\chi_F(x)={\mc D}_F^{\ga}f(x)$.  Then $\int_a^bh(x)d_F^{\ga}x=f(b)-f(a)$.
\item
(Integration by parts)  Assume:  (i)  $u$ is continuous on $[a,b]$ and $Sch(u)\subset F$.
(ii) ${\mc D}_F^{\ga}u(x)$ exists and is F-continuous on $[a,b]$.  (iii)  v is F-continuous
on $[a,b]$.  Then
\bq\label{511}
\int_a^buvd_F^{\ga}x=\left[u(x)\int_a^xv(x')d_F^{\ga}x']\right|_a^b-\int_a^b{\mc D}_F^{\ga}
u(x)\int_a^xv(x')d_F^{\ga}x'd_F^{\ga}x
\end{equation}
\end{enumerate}
\indent
Some examples are given relative to applications and we mention e.g.
\begin{example}
Following \cite{k23} one has a local fractal diffusion equation 
\bq\label{512}
{\mc D}_{F,t}^{\ga}(W(x,t))=\frac{\chi_F(t)}{2}\frac{\pp^2}{\pp x^2}W(x,t)
\end{equation} 
with solution 
\bq\label{5.13}
W(x,t)=\frac{1}{(2\pi S_F^{\ga}(t))^{1/2}}exp\left(\frac{-x^2}{2S_F^{\ga}(t)}\right)
\end{equation}
\end{example}
The appendix to \cite{p3} also gives some formulas for repeated integration and
differentiation.  For example it is shown that
\bq\label{513}
({\mc D}_F^{\ga})^2(S_F^{\ga}(x))^2=2\chi_F(x);\,\,\int_a^{x'}(S_F^{\ga}(x))^nd^{\ga}_Fx=
\frac{1}{n+1}(S_F^{\ga}(x'))^{n+1}
\end{equation}
We refer to \cite{k23,p3} for much other interesting stuff.$\hfill\bs$

\subsection{POSSIBLE APPLICATIONS}

Example 5.2 gives a model of a local fractal diffusion equation where the transition
times $t\in F$ are rare (or infrequent) and this was developed also in \cite{k23} for
local fractional derivatives in a broader context.
It is not clear whether or not the fractal calculus above
will be useful in describing quantum paths of Hausdorff dimension two for example or
other fractal curves such as Peano or von Koch curves (for which \cite{k23} looks more
promising - see also \cite{a10,c13,c31});
such curves are not of the form 
$y(t)$ with $t\in F$ for sets F of zero topological dimension (such as Cantor sets). 
Nevertheless for paths with rare transitions we indicate
here heuristically how this fractal calculus from \cite{p3} might be applied to
the SE in the spirit of Nottale et al as sketched in Examples 2.6 and 2.7 along with Remark
2.1.  Thus instead of a continuous nondifferentiable curve think of an F-continuous path
$y(t)$ based on 
$t\in F\subset [a,b]$ where F is an $\ga$-perfect fractal set (e.g. a Cantor set) of
H-dimension $\ga$ (assume here that $y$ is actually defined on all $[a,b]$). If $y(t)$
changes direction (e.g. increasing to decreasing) at
$t_0$ then one expects different left and right side derivatives ${\mc D}_F^{\ga}y(t_0)$.
Thus we define
\bq\label{515}
{}^{+}{\mc D}_F^{\ga}y(t_0)=F-limit_{t\to t_0^{+}}\frac{y(t)-y(t_0)}{S_F^{\ga}(t)
-S_F^{\ga}(t_0)};
\end{equation}
$${}^{-}{\mc D}_F^{\ga}y(t_0)=F-limit_{t\to t_0^{-}}\frac
{y(t)-y(t_0)}{S_F^{\ga}(t)-S_F^{\ga}(t_0)}$$
Then as in Example 2.6 for example we write ${\bf (D23)}\,\,b_{\pm}(t)={}^{\pm}{\mc
D}_F^{\ga}y(t)$ and set ${\bf (D24)}\,\,V=(1/2)(b_{+}+b_{-})$ with
$U=(1/2)(b_{+}-b_{-})$ with ${\mc V}=V-iU$.  Now for functions $f(y(t),t)$ which are suitably
smooth in the y variable one wants to differentiate in t.  We recall that for F-limits as in
\eqref{515} $t\in F$ along with 
$t_0$.  The set F could perhaps be regarded as $Sch(y)$ so $y(t)$ is characterized by
changes on F with no specified behavior otherwise.  
\begin{proposition}
Let $f$ be smooth and $y(t)$ continuous on $[a,b]\supset F$ and ${\mc D}^{\ga}_Fy(t)=0$ for
$y\ne F$ as  specified in \eqref{59}.  Then
\bq\label{516}
{\mc D}_F^{\ga}f(y(t))=[{\mc D}_F^{\ga}y(t)]f'(y(t))\,\,\,(t\in F)
\end{equation}
{\it Proof}:  One has $f(s)-f(\gs)=f'(c)(s-\gs)$ where $\gs\leq c\leq s$.  Now write
\bq\label{517}
\frac{f(y(t'))-f(y(t))}{S_F^{\ga}(t')-S_F^{\ga}(t)}=\frac{(y(t')-y(t))f'(c)}{S_F^{\ga}(t')
-S_F^{\ga}(t)}\to {\mc D}_F^{\ga}y(t)f'(y(t))
\end{equation}
for $t\in F$ (with limit zero otherwise).  Here $y(t)\leq c\leq y(t')$ and $y$ is assumed
continuous which squeezes c.  {\bf QED}
\end{proposition}
\indent
Instead of assuming one sided derivatives
everywhere on $[a,b]$ as in Section 2 we suppose only one sided $\ga$-derivatives on F as in
\eqref{515}.  The undercurrent here is to imagine a fractal space time with some average
dimension 3 or 4 say, as in the unstructured ${\mf E}^{\infty}$ of El Naschie.  Thus
we might be able to ``see" a continuous path but in fact the underlying
phenomena could involve points $(t,y(t))$ for $t\in F$ as a fractal graph (whose dimension
is basically irrelevant at the moment).  One could imagine a geodesic having an
underlying fractal structure, no matter what we may fuzzily see in an ambient continuum. 
Hence one could work with 
$b_{\pm}$ and $(U,V)$ as in {\bf (D23)}-{\bf (D24)} and think of some Brownian fluctuations 
(suitably appended) in addition, as in \eqref{2.17}-\eqref{2.18}.  The eventual ``geodesics"
in the microstructure involve derivatives ${\mc D}_F^{\ga}$ in the t variable and
consequently would be tied to F intrinsically.  Away from F there is no microstructure but we
may see something in the continuum by visually blending together points.
\\[3mm]\indent
Thus we introduce a Wiener process $d\xi_i$ as in Example 2.6 with ${\bf
(D25)}\,\,<d\xi_{+}^2>=2Ddt=-<d\xi_{-}^2>$ (where $D=\hbar/2m$).  Although $dt$ makes no
sense in F we know ${}^{\pm}{\mc D}_F^{\ga}y(t)=0$ for $t\ne F$ so one can envision 
equations as in \cite{k23} involving ${}^{\pm}{\mc D}_F^{\ga}y(t)$
provided $\chi_F(t)$ is inserted at appropriate places (i.e. there will only be evolution for
$t\in F$).  First as in \cite{c7}, following \cite{k23,n6,n5}, one writes heuristically
for 1-D and $f=f(t,y(t))$ smooth in the $y$ variable
\bq\label{518}
\frac{d_{+}f}{dt}=\chi_F\pp_t+b_{+}(t)\pp_yf+\chi_FD\pp_y^2f;\,\,\frac{d_{-}f}{dt}=
\chi_F\pp_tf+b_{-}\pp_yf-\chi_FD\pp_y^2f
\end{equation}
\indent
{\bf REMARK 5.1.}
There are a number of papers about diffusion and fractional diffusion on fractals (see e.g.
\cite{a5,b55,g3,r1}) and some anomalous features arise.  We omit discussion of this for the
moment.
$\hfill\bs$
\\[3mm]\indent
{\bf REMARK 5.2.}
Proceeding now from \eqref{518} we now obtain (using {\bf (D23)}-{\bf (D24)} the formula
${\bf (D26)}\,\,d'/dt=\chi_F\pp_t-iD\chi_F\pp_y^2+{\mc V}\pp_y$ where of course ${\mc V}=
V-iU=0$ for $t\ne F$.  
Now the geodesic equation for a free ``particle" is from {\bf (A22)}
${\bf (D27)}\,\,m(d'/dt){\mc V}=0$.  Here we note that ${\mc V}={\mc
V}(y(t))$ must be assumed to be of the form $f(y(t),t)$ smooth in some sense in y and 
differentiable in some sense in t.
This problem
seems to arise also in the derivations of the SE in \cite{n5} for example since $\pp_tU$ and
$\pp_tV$ are not a priori well defined and this is one reason we include the present
discussion.  It would be interesting to pursue it further here and in the context of
\cite{n5}. 
Further since e.g.
$V=(1/2)(b_{+}+b_{-})$ with
$b_{\pm}(t)= {}^{\pm}{\mc D}_F^{\ga}y(t)$ there is even no assurance that ${\mc
D}_F^{\ga}b_{+}(t)$ will be defined (let alone one sided derivatives $b_{\pm}$ as in
\cite{n5} and Section 2 or {\bf (D23)}).  We will examine this now in more detail
and refer to \cite{a10,c13,c31} for variations and refinements.  The derivation of
Nottale given here and in Example 2.5, although heuristic, is however more revealing
in showingjust how the formal U and V play a role in creating the SE (and a 
quantum potential).
Thus from \cite{n5} (cf. also \cite{c7,n6}) one defines $d_{\pm}$ as in \eqref{518} (with
one sided derivatives $d_{\pm}$) and sets ${\bf (D28)}\,\,d_v/dt=(1/2)(d_{+}+d_{-})/dt$ with
$d_u/dt=(1/2)(d_{+}-d_{-})/dt$ so that ${\bf (D29)}\,\,d'\sim d_v-id_u$ agrees with the
formula {\bf (D26)}.  Then ${\bf (D30)}\,\,d'{\mc V}=(d_v-id_u)(V-iU)=(d_uV-d_uU)-i(d_uV+
d_vU)$.  Given no driving force both real and imaginary parts of the complex acceleration
$d'{\mc V}/dt$ vanish and we note that in \cite{n6} one defines acceleration via $\ddot{y}=
(1/2)(d_{+}d_{-}+d_{-}d_{+})y/dt^2$ which is exactly the real part of $d'{\mc V}/dt$, namely
${\bf (D31)}\,\,d_vV-d_uU$.  Hence in our situation involving ${}^{\pm}{\mc D}_F^{\ga}$ one
has for the real part of {\bf (D27)} $(d'/dt){\mc V}=0$ the expression
\bq\label{519}
({}^{+}{\mc D}_F^{\ga}{}^{-}{\mc D}_F^{\ga}+{}^{-}{\mc D}_F^{\ga}{}^{+}{\mc D}_F^{\ga})
y(t)=0\,\,\,(t\in F)
\end{equation}
which requires of course that this combination of derivatives makes sense.  For the imaginary
part of
$(d'/dt){\mc V}=0$ we have ${\bf (D32)}\,\,(1/2)(d_{+}^2-d_{-}^2)y=0$ so in our situation we
require
${\bf (D33)}\,\,[({}^{+}{\mc D}_F^{\ga})^2-({}^{-}{\mc D}_F^{\ga})^2]y(t)=0\,\,\,(t\in F)$
which would be automatic if ${}^{-}{\mc D}_F^{\ga}y={}^{+}{\mc D}_F^{\ga}y={\mc D}_F^{\ga}y$.
\\[3mm]\indent
Now using \eqref{518} one has expressions for $d_{\pm}$ acting on suitable functions
$f(y(t),t)$ and the transition to the SE involves using ${\bf (D34)}\,\,(d_v/dt)=
\pp_t+V\pp$ and $(d_u/dt)=D\pp^2+U\pp$ which are then applied to $U,V$ considered as
functions of $(y(t),t)$.  Hence one must assume here that e.g. $V(y(t),t)$ is smooth in
y and differentiable in the second variable t, while $y(t)$ remains fractal in nature.
Now let $\rho(y(t),t)$ be the probability density of $y(t)$ in a Wiener process context.
Classically one knows there are then forward and backward Fokker-Planck (FP) equations
\bq\label{520}
\pp_t\rho+div(\rho b_{+})=D\gD\rho;\,\,\pp_t\rho +div(\rho b_{-})=-D\gD\rho
\end{equation}
and by adding and subtracting one gets ${\bf (D35)}\,\,\pp_t\rho+div(\rho V)=0$ and $div(\rho
U)-D\gD
\rho=0$ (here $U,V$ are essentially regarded as statistical averages $<u>$ and $<v>$).
Hence let us assume a similar situation prevails in our case and stipulate that
${\bf (D36)}\,\,U=D\pp log(\rho)$ in the 1-D fractal situation.  Now using {\bf (D34)}
the imaginary part of the acceleration satisfies ${\bf (D37)}\,\,d_uV+d_vU=0=\pp_tU+\pp (UV)
+D\pp^2V$ so $\pp_tU=-D\pp^2V-\pp(UV)$.  One can also assume $V=2D\pp S$ via Lagrangian
arguments which leads to $\psi=\sqrt{\rho}exp(iS)$ and thence easily to the SE
(cf. \cite{c7,n5}).  In fact we can simply write ${\bf (D38)}\,\,2iD(d'/dt)(\pp log(\psi))=0$
where ${\bf (D39)}\,\,log(\psi)=(1/2)log(\rho)+iS$ so $iD\pp log(\psi)=iD\pp log(\rho)
-2D\pp S=(V-iU)={\mc V}$.  To produce the SE use then the expression ${\bf (D40)}\,\,
(d'/dt)=(\pp_t-iD\pp^2)+{\mc V}\pp=(d_v-id_u)/dt$ to get (using {\bf (D39)})
\bq\label{521}
0=2iDm[\pp_t\pp log(\psi)-iD\pp^2\pp log(\psi)-2iD(\pp log(\psi)\pp)(\pp log(\psi)]
\end{equation}
leading to the SE ${\bf (D41)}\,\,i\hbar\pp_t\psi+(\hbar^2/2m)\pp^2\psi=0$ (as a geodesic
equation - recall we are taking $D=\hbar/2m$).$\hfill\bs$
\\[3mm]\indent
In the present situation the meaning of $\rho$ as a probability density can surely be
retained (on F).  It would be nice if one could rephrase the assumptions of $U,V,\rho$, 
as functions of $(y(t),t)$, being smooth in y and differentiable in the second variable t,
and reduce the whole context to behavior on $t\in F$ and $(y(t),t)\in P$ where P is some
kind of ``quantum" path.  To use the calculus of \cite{p3}
however for $(U,V,\rho)(y(t),t)$ we would need to further develop concepts like the chain rule
(cf. Proposition 5.1) and perhaps various equalities would have to be replaced by
inequalities. Thus for now we will assume $y(t)$ is continuous for $t\in [a,b]$ and treat
$\pp_y\sim\pp$ symbolically while $\pp_t$ could naturally mean ${\mc D}_F^{\ga}$ in the t
variable written as $\pp_F^{\ga}$.  The symbolic $\pp$ could perhaps eventually reduce at
times to ${\mc D}_G^{\gb}\sim\pp_G^{\gb}$ in $y$ where $(y(t),t)\in P$ implies that $y(t)\in
G$ for
$t\in F$ where G is some fractal set of Hausdorff dimension $\gb$.  Then formally repeating
the above calculations one arrives heuristically at a SE
\bq\label{522}
i\hbar\pp_F^{\ga}\psi=-\frac{\hbar^2}{2m}\chi_F\pp^2\psi;\,\,\psi=\sqrt{\rho}e^{iS};\,\,
U=\frac{\hbar}{2m}\pp log(\rho);\,\,V=\frac{\hbar}{m}\pp S
\end{equation}
\indent
{\bf REMARK 5.3.}
One notes that (with the assumptions above) the SE in $\psi$ does not explicitly involve the
fractal effect described by U.  However the corresponding HJ equation (cf. Example 2.6)
\bq\label{523}
\pp_F^{\ga}S+\chi_F\frac{(\pp S)^2}{2m}+Q=0;\,\,Q=-\frac{m}{2}U^2-\frac{\hbar}{2}\pp U
\end{equation}
reveals the effect of U.  Here ${\bf (D42)}\,\,U=(1/2)(b_{+}-b_{-})=(1/2)({}^{+}{\mc
D}_F^{\ga}y(t)-{}^{-}{\mc D}_F^{\ga}y(t))$.
We remark also that the above procedure illustrates again the strong connection between 
diffusion and the SE (cf. \cite{n9,n15,n6}).
$\hfill\bs$
\\[3mm]\indent
{\bf REMARK 5.4.}
The idea of having a fractal microstructure for the vacuum is of course reminiscent of the
ether idea about which there already seems to be a substantial amount of recent material
(cf. \cite{d29,j3,s21,s22,s23,v3,v4}).  We will return to this at another time in noting
that explicit modifications to the SE are indicated in \cite{s22} (cf. also \cite{s21}
regarding the Bohmian theory).  It appears also that the viewpoints and information arising
from \cite{v3,v4} should be of critical importance in the future development of quantum
theory and cosmology.
$\hfill\bs$

\end{document}